\documentclass[final,3p,times,fleqn]{elsarticle}
\usepackage{mathtools,booktabs}
\usepackage{bm}
\usepackage{autobreak}
\usepackage{amsmath}
\usepackage{amsthm}
\usepackage{amssymb}
\usepackage{siunitx}
\usepackage{gensymb}
\usepackage{subfigure}
\usepackage{epstopdf}
\usepackage{color}
\usepackage{hyperref}
\usepackage{numprint}
\usepackage{CJKutf8}
\newtheorem{remark}{Remark}
\biboptions{sort&compress}
\begin{document}
\begin{CJK*}{UTF8}{gbsn}
\begin{frontmatter}
	\title{A ternary phase-field model for two-phase flows in complex geometries}
	\author[a,b,c]{Chengjie Zhan}
	\author[a,b,c]{Zhenhua Chai \corref{cor1}}
	\ead{hustczh@hust.edu.cn}
	\author[a,b,c]{Baochang Shi}
	\address[a]{School of Mathematics and Statistics, Huazhong University of Science and Technology, Wuhan 430074, China}
	\address[b]{Institute of Interdisciplinary Research for Mathematics and Applied Science, Huazhong University of Science and Technology, Wuhan 430074, China}
	\address[c]{Hubei Key Laboratory of Engineering Modeling and Scientific Computing, Huazhong University of Science and Technology, Wuhan 430074, China}
	\cortext[cor1]{Corresponding author.}
\begin{abstract} 
	In this work, a ternary phase-field model for two-phase flows in complex geometries is proposed. In this model, one of the three components in the classical ternary Cahn-Hilliard model is considered as the solid phase, and only one Cahn-Hilliard equation with degenerate mobility needs to be solved due to the condition of volume conservation, which is consistent with the standard phase-field model with a single-scalar variable for two-phase flows. To depict different wetting properties at the complex fluid-solid boundaries, the spreading parameters in ternary phase-field model are determined based on the Young's law, in which the liquid-solid surface tension coefficient is assumed to be a linear function of gas-liquid surface tension coefficient and related to the contact angle and the minimum curvature of the solid surface. In addition, to achieve a high viscosity in the solid phase and preserve the velocity boundary conditions on the solid surface, the phase-field variable of the solid phase is also used to derive the modified Navier-Stokes equations. To test the present model, we further develop a consistent and conservative Hermite-moment based lattice Boltzmann method where an adjustable scale factor is introduced to improve the numerical stability, and conduct the numerical simulations of several benchmark problems. The results illustrate that present model has the good capability in the study of the two-phase flows in complex geometries.
\end{abstract}
\end{frontmatter}
\end{CJK*}	

\section{Introduction}
The two-phase flows in complex geometries are usually observed in both nature and engineering, and also become the fundamental problems in the fields of the enhanced oil recovery \cite{Alvarado2010,Maghzi2012ETFS}, the geological CO$_2$ sequestration \cite{Fan2019Energy}, the advanced manufacture \cite{Korneev2020CAD,Blakey2021MD}, and so on. Due to the difficulty in exploring the physical mechanism and the expensive cost of experimental approach, the numerical simulation has become an effective method in the study of such complex problems \cite{Abriola1989EHP}.
To describe two immiscible fluid phases, the Cahn-Hilliard (CH) equation \cite{Cahn1996EJAM}, as one of the popular used phase-field model \cite{Badalassi2003JCP,Jacqmin1999JCP}, has been widely applied due to its properties of the mass conservation and thermodynamic consistency \cite{Abels2013}. 

When we consider the fluid flows in complex geometries, how to implement the boundary conditions is usually a tough issue. To solve this problem, some approaches have been proposed, including the immersed boundary method \cite{Peskin1977JCP,Fogelson1988JCP}, the fictitious domain method \cite{Angot1999MMAS,Ramiere2007CMAME}, the volume penalization method \cite{Angot1999NM,Kolomenskiy2009JCP}, the smoothed profile method \cite{Nakayama2005PRE,Nakayama2008EPJE}, and the diffuse domain method \cite{Li2009CMS,Lervag2015CMS}. In these methods, a larger and regular domain containing the original complex geometries and the fluid-solid boundaries is first introduced, and simultaneously a sharp/diffuse indicator function at Cartesian/Lagrangian grids is used to distinguish the interface between fluid and solid phases. Then the boundary conditions at fluid-solid interface are treated as the source terms and/or external force terms into the reformulated governing equations or numerical schemes, and thus the difficulty in directly treating the complex boundary conditions can be avoided. 

For the multiphase problems, the wetting property also has an obvious influence on the fluid dynamics. However, the application of the wetting boundary condition usually needs to use the local normal vector of the solid boundaries, which is difficult to be determined for the complex geometries. Up to now, some of the above-mentioned methods \cite{Horgue2014CF,Patel2018JCP,Shahmardi2021JCP,Sharaborin2021,Aland2010CMES,Guo2021JFM,Yang2023JCP} have been successfully applied to investigate the multiphase flows with wetting boundary conditions, where a smooth indicator function for the solid phase is naturally combined with the diffuse-interface methods for two-phase flows. However, in these methods, how to reconstruct the governing equations or numerical schemes is also an important issue, and some additional treatments are needed.  

From another point of view, the two-phase flows in complex geometries can be seen as the three-component flows in a larger and regular domain, and one of which is considered as the solid phase. Actually, in the phase-field method, Boyer and Lapuerta \cite{Boyer2006MMNA} designed a consistent free energy for the ternary CH model, which can reduce to the standard form for two-phase cases when one phase disappears. Although there have been many phase-field models for multiphase flows \cite{Kim2007CMAME,Boyer2014MMMAS,Dong2018JCP} up to now, we only focus on this ternary-fluid model in present work. 
Based on the model in Ref. \cite{Boyer2006MMNA}, Yi \cite{Yi2020FDR} proposed a phase-field method for the moving solid object in two-phase flows, where the rigidity constraint of the solid phase and the no-slip boundary condition on the solid-fluid interface is imposed by attributing a high viscosity to the solid phase. This work aims to handle complex two-phase flows with a moving solid object, but unfortunately, their results show that the solid phase has some deformation. What is more, to consider a hydrophilic or hydrophobic solid object, the surface tension coefficients of the gas/liquid/solid interfaces are artificially given without any discussion. 
Later, Rohde and Wolff \cite{Rohde2021MMMAS} presented a ternary phase-field model for two-phase flow with precipitation and dissolution. In their model, when they consider the wetting boundary condition, the surface tension coefficients related to the solid phase are assumed to be much larger than the surface tension coefficient between two fluid phases, such that the mechanical balance condition at the triple points can be preserved as far as possible. Additionally, to avoid the restriction of the maximum principle for fourth-order CH equation, the free energy is designed to enforce a relaxed constraint $\left(-\delta,1+\delta\right)$ for the order parameter by using an unbounded potential function, instead of the original range restriction $\left[0,1\right]$. Furthermore, the Navier-Stokes (NS) equations for the fluid flows in their model are also more reasonable through considering some conditions of consistency. However, in their work, the values of the surface tension coefficients are not given, and the capacity in predicting contact angle has also not been tested.  
More recently, Panter et al. \cite{Panter2023CP} used a ternary phase-field model \cite{Boyer2006MMNA} to investigate the capillary rise within rough structures, in which one of three order parameters is defined as the rough structures to smooth and curve the solids. However, in their model, the wetting property of the rough structures seems not to be considered, and the solid surface at the base of the pillars is still treated via a boundary condition, rather than a diffuse interface.

In this work, to overcome the problems mentioned in above literature, we present some improvements and propose a mathematical model for two-phase flows in complex geometries. In the present model, a specific expression of the liquid-solid surface tension coefficient is first designed, which is more accurate in predicting the contact angles. Then a degenerate mobility is used to overcome the mass leakage among different phases in the CH model \cite{Yue2007JCP}. Finally, the consistent and conservative NS equations is adopted to describe the fluid flows, and the volume fraction of the solid phase is used to achieve the high viscosity in solid phase and ensure the velocity boundary conditions on solid interface. To numerically solve the present model, the lattice Boltzmann method \cite{Higuera1989EL,Chen1998ARFM,Aidun2010ARFM} is applied, which is an efficient numerical tool in the simulation of complex fluid systems, especially for the multiphase flows due to the features of the easy implementation of boundary conditions and fully parallel algorithm \cite{Chen1998ARFM}.    

The rest of this paper is organized as follows. In Section \ref{model}, the original three-component CH model is first introduced, and then the model for two-phase flows in complex geometries and the relation of surface tension coefficients are designed, followed by the consistent and conservative NS equations. In Section \ref{LBM}, the consistent and conservative LB method for the proposed phase-field model is developed. In Section \ref{Numerical}, the numerical validations are performed by several benchmark tests, and the problems of displacement in a complex channel and a porous medium are studied. Finally, some conclusions are summarized in Section \ref{Conclusions}.

\section{Ternary phase-field model for two-phase flows in complex geometries}\label{model}
\subsection{The Cahn-Hilliard model for three-component flows}\label{CH3f}
The mix free energy for three-component flows can be expressed as \cite{Boyer2006MMNA} 
\begin{equation}\label{eq-freeEnergy}
	E=\int_{\Omega}f_b\left(\phi_0,\phi_1,\phi_2\right)+\frac{3D}{8}\sum_{p=0}^{2}\gamma_p|\nabla\phi_p|^2\mathrm{d}\Omega,
\end{equation}
where $\phi_p$ is the phase-field variable of phase $p$ satisfying $\phi_0+\phi_1+\phi_2=1$, $D$ is the interface thickness, and the spreading coefficient $\gamma_p$ is related to the pairwise symmetric surface tension coefficient $\sigma_{pq}$,
\begin{equation}
	\begin{aligned}
		\gamma_0&=\sigma_{01}+\sigma_{02}-\sigma_{12},\\
		\gamma_1&=\sigma_{01}+\sigma_{12}-\sigma_{02},\\
		\gamma_2&=\sigma_{02}+\sigma_{12}-\sigma_{01}.
	\end{aligned}
\end{equation}
$f_b\left(\phi_0,\phi_1,\phi_2\right)$ is the bulk free energy density, and is given by \cite{Boyer2006MMNA}
\begin{equation}
	f_b\left(\phi_0,\phi_1,\phi_2\right)=\frac{12}{D}\left[\frac{\gamma_0}{2}\phi_0^2\left(1-\phi_0\right)^2+\frac{\gamma_1}{2}\phi_1^2\left(1-\phi_1\right)^2+\frac{\gamma_2}{2}\phi_2^2\left(1-\phi_2\right)^2\right].
\end{equation}
The chemical potential of $p$-th fluid can be calculated by the variation of mix free energy,
\begin{equation}
	\mu_p=\frac{\delta E}{\delta \phi_p}=\frac{\partial f_b}{\partial\phi_p}-\frac{3D}{4}\gamma_p\nabla^2\phi_p.
\end{equation}

Then one can obtain the following CH equation through minimizing the mix free energy,
\begin{equation}
	\frac{\partial\phi_p}{\partial t}+\nabla\cdot\left(\phi_p\mathbf{u}\right)=\nabla\cdot M_p\nabla\bar{\mu}_p,\quad p=0,1,2,
\end{equation}
where $\mathbf{u}$ is the fluid velocity, $M_p$ is the mobility, $\bar{\mu}_p=\mu_p+L$ with $L$ being a Lagrangian multiplier used to ensure the volume conservation of phase-field variables.	
To determine the expression of $L$, one can assume $M=M_0\gamma_0=M_1\gamma_1=M_2\gamma_2$ and sum all CH equations with $p=0,1,2$,
\begin{equation}\label{eq-sumCH}
	0=\frac{\partial 1}{\partial t}+\nabla\cdot\left(1\mathbf{u}\right)=\nabla\cdot M\sum_{p=0}^2\left(\frac{\mu_p}{\gamma_p}+\frac{L}{\gamma_p}\right),
\end{equation}
where the conservative condition of $\phi_0+\phi_1+\phi_2=1$ is used.
From the above equation, $L$ can be given by
\begin{equation}
	L=-\frac{\gamma^T}{3}\sum_{q=0}^2\frac{\mu_q}{\gamma_q},\quad\frac{3}{\gamma^T}=\frac{1}{\gamma_0}+\frac{1}{\gamma_1}+\frac{1}{\gamma_2}.
\end{equation}
With the above expression of $L$, the CH model can be rewritten as
\begin{subequations}\label{eq-CHmodel}
	\begin{equation}
		\frac{\partial\phi_p}{\partial t}+\nabla\cdot\left(\phi_p\mathbf{u}\right)=\nabla\cdot\left(\frac{M}{\gamma_p}\nabla\bar{\mu}_p\right),
	\end{equation}
	\begin{equation}
		\bar{\mu}_p=\frac{\gamma^T}{3}\sum_{q\neq p}\frac{1}{\gamma_q}\left(\mu_p-\mu_q\right),\quad p=0,1,2.
	\end{equation}
\end{subequations}

\subsection{The Cahn-Hilliard model for two-phase flow in complex geometries}\label{CH2f1s}
For the two-phase flows in complex geometries, the system can be viewed as a three-phase case and be governed by the three-component CH model in Section \ref{CH3f} with some extra limitations. In this work, we set phase 0 to be the solid phase, and thus $\phi_0$ becomes a smoothed indicator function to identify the solid phase ($\phi_0=1$) and fluid phase ($\phi_0=0$). 

\begin{figure}
	\centering
	\includegraphics[width=2.5in]{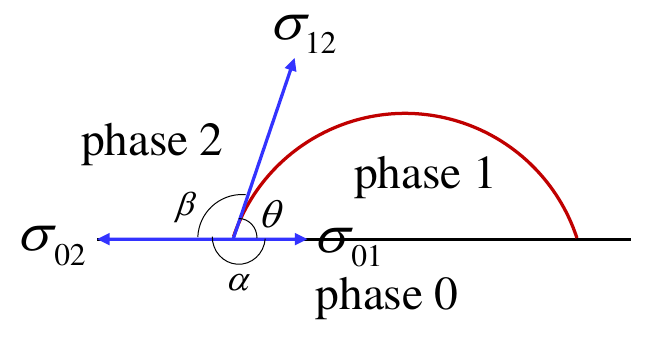}
	\caption{Three-phase contact diagram.}
	\label{fig-sigma}
\end{figure}

To include the effect of the wetting property on the solid surface in a three-phase system, the Young's law \cite{Young1805} can be applied, i.e., $\sigma_{02}=\sigma_{01}+\sigma_{12}\cos\theta$ where $\theta$ is the contact angle of fluid phase 1 to the solid phase 0 (see the schematic in Fig. \ref{fig-sigma}).  
Additionally, following the work of Rohde and Wolff \cite{Rohde2021MMMAS}, the mechanical balance of surface tension coefficients and the angles at the triple points can be given by 
\begin{equation}\label{eq-triple}
	\frac{\sin\alpha}{\sigma_{12}}=\frac{\sin\beta}{\sigma_{01}}=\frac{\sin\theta}{\sigma_{02}},
\end{equation}
where $\alpha+\beta+\theta=2\pi$. When the phase 0 is considered as solid phase, we need to set $\alpha=\pi$, and $\sin\alpha=0$. Thus, the surface tension coefficients need to satisfy $\sigma_{12}\ll\sigma_{01},\sigma_{02}$ \cite{Rohde2021MMMAS}. Actually, under this condition, the terms with large value of $\gamma_0$ in the mix free energy (\ref{eq-freeEnergy}) can be seen as the penalties for solid phase 0, and there should be an optimal penalizing parameter to give the most reasonable result. However, as far as we know, there are no specific expressions of the spreading coefficients or fluid/solid surface tension coefficients at present.
For this reason, in this work, we assume $\sigma_{01}=k\sigma_{12}$ and give the following expression of $k$, 
\begin{equation}\label{eq-kk}
	k=\frac{9}{1+\kappa}\left[1+2\mathrm{sgn}\left(\theta-\frac{\pi}{2}\right)\sin^2\left(\frac{\pi}{4}\cos\theta\right)\right],
\end{equation}
which is a characteristic parameter related to the curvature and the contact angle of the solid surface. Here $\kappa$ is the minimum dimensionless curvature of a solid structure and $\mathrm{sgn}\left(x\right)$ is a sign function. Specially, $\kappa=0$ denotes the case of an ideal solid plate.
With the above formula, the parameter $\gamma_p$ can be rewritten as 
\begin{equation}
	\begin{aligned}
		\gamma_0&=\left(2k+\cos\theta-1\right)\sigma_{12},\\
		\gamma_1&=\left(1-\cos\theta\right)\sigma_{12},\\
		\gamma_2&=\left(1+\cos\theta\right)\sigma_{12}.
	\end{aligned}
\end{equation}

Moreover, to eliminate the defects of the CH equation in terms of the boundedness and phase mass conservation, a degenerate mobility $M=M_{cst}+M_{deg}f\left(\bm{\phi}\right)$ is applied in this work, where $M_{cst}$ and $M_{deg}$ are two positive constants and $f\left(\bm{\phi}\right)$ is a function with $\bm{\phi}=\left(\phi_0,\phi_1,\phi_2\right)$. Here it should be noted that the form of mobility does not influence the formal derivation of the present model.

\begin{remark}
	Compared to the original CH model for three-phase flows, the present model for two-phase flows in complex geometries needs to set the values of $\sigma_{01}$ and $\sigma_{02}$ according to Eq. (\ref{eq-kk}) and Young's law. 
\end{remark}
\begin{remark}
	For the two phase flows in stationary geometries, only $\phi_1$ governed by the CH model (\ref{eq-CHmodel}) needs to be solved and $\phi_2$ can be calculated by the conservation of volume, which is consistent with the standard phase-field equation with a single-scalar variable for two-phase flows.
\end{remark}
\begin{remark}
	The phase-field variable $\phi_0$ for solid phase is usually not governed by the CH model (\ref{eq-CHmodel}) with $p=0$, and it does not need to be evolved for stationary geometries. In addition, the evolution of $\phi_0$ needs to be included when the problems including solid moving, deformation or phase transition are considered, for instance, the phase-field equations for dendritic growth \cite{Karma2001PRL,Echebarria2004PRE}. 
\end{remark}

\subsection{The consistent and conservative Navier-Stokes equations for flow field}
To depict the fluid flows, the following consistent and conservative NS equations are adopted,
\begin{subequations}\label{eq-NS}
	\begin{equation}
		\nabla\cdot\mathbf{u}=0,
	\end{equation}
	\begin{equation}\label{eq-NS2}
		\frac{\partial\left(\rho\mathbf{u}\right)}{\partial t}+\nabla\cdot\left[\left(\rho\mathbf{u}-\mathbf{S}\right)\mathbf{u}\right]=-\nabla P+\nabla\cdot\frac{\mu}{1-\phi_0}\left[\nabla\mathbf{u}+\left(\nabla\mathbf{u}\right)^\top\right]+\mathbf{F}_s+\phi_0\rho\mathbf{f}+\mathbf{F}_b,
	\end{equation}
\end{subequations}
where $\rho$ is the density, $P$ is the pressure. $\mu/\left(1-\phi_0\right)$ is the effective dynamic viscosity with $\left(1-\phi_0\right)$ being used to give a high viscosity in solid phase and a normal viscosity in fluid region with $\phi_0=0$. $\mathbf{S}$ is the extra mass flux across the interface of two fluid phases, and is given by $\mathbf{S}=\sum_{p=0}^2\rho_pM\nabla\bar{\mu}_p/\gamma_p$ for the CH equation (\ref{eq-CHmodel}). $\mathbf{F}_s=\sum_{p=0}^2\mu_p\nabla\phi_p$ is the surface tension force, and $\mathbf{F}_b$ represents the body force, such as gravity. The velocity boundary condition is imposed by the force $\phi_0\rho\mathbf{f}$, and the form of fluid-structure interaction $\mathbf{f}$ will be stated in the next section. 

In addition, the density and dynamic viscosity are usually set as the linear functions of the phase-field variables,
\begin{equation}
	\rho=\sum_{p=0}^2\rho_p\phi_p,\quad \mu=\sum_{p=0}^2\mu_p\phi_p,
\end{equation}
where $\rho_p$ and $\mu_p$ are the density and viscosity of the pure $p$-th fluid. 

\section{Lattice Boltzmann method for the three-component phase-field model}\label{LBM}
In this section, we will develop the LB method to solve the mathematical model of three-component flows under the general frameworks \cite{Chai2020PRE,Chai2023PRE}. However, it should be noted that the degenerate mobility in CH equation is a function of the phase-field variables and may become zero, which may cause the numerical instability in the standard LB model. To overcome this problem, we rewrite the CH equation into the following form with a constant diffusion coefficient,
\begin{equation}\label{eq-CH}
	\frac{\partial\phi_p}{\partial t}+\nabla\cdot\left(\phi_p\mathbf{u}\right)=\nabla\cdot \frac{M_{cst}+M_{deg}}{\gamma_p}\left[\nabla\bar{\mu}_p-M_{deg}\frac{1-f\left(\bm{\phi}\right)}{M_{cst}+M_{deg}}\nabla\bar{\mu}_p\right].
\end{equation}
For the traditional three-phase flows, the so-called \emph{3f} model composed of the equations with $p=0$ and 1 needs to be solved, while in the \emph{2f1s} model for two-phase flows in complex geometries, only the CH equation with $p=1$ needs to be considered. 

In the LB method, the evolution equation for CH equation (\ref{eq-CH}) can be given by 
\begin{equation}
	f_i^p\left(\mathbf{x}+\mathbf{c}_i\Delta t,t+\Delta t\right)=f_i^p\left(\mathbf{x},t\right)-\Lambda_{ij}^p\left[f_j^p\left(\mathbf{x},t\right)-f_j^{p,eq}\left(\mathbf{x},t\right)\right]+\Delta t\left(\delta_{ij}-\frac{\Lambda_{ij}^p}{2}\right)R_j^p,
\end{equation}  
where $f_i^p\left(\mathbf{x},t\right)$ is the distribution function of phase-field variable $\phi_p$ in the direction of $i$ ($i=0,1,\cdots,Q-1$ with $Q$ being the number of discrete velocity directions) at position $\mathbf{x}$ and time $t$. $\mathbf{c}_i$ is the discrete velocity, $\Delta t$ is the time step, and $(\Lambda_{ij}^p)$ represents the collision matrix. In addition, the distribution functions $f_i^{p,eq}$ and $R_i^p$ are designed as
\begin{equation}
	f_i^{p,eq}=\begin{cases}
		\phi_p+\left(\omega_{i,p}-1\right)\eta_p\bar{\mu}_p, &i=0,\\
		\omega_{i,p}\eta_p\bar{\mu}_p+\omega_i\mathbf{c}_i\cdot\phi_p\mathbf{u}/c_{s,p}^2, &i\neq 0,
	\end{cases}\quad R_i^p=\omega_{i,p}\left[\frac{\mathbf{c}_i\cdot\partial_t\left(\phi_p\mathbf{u}\right)}{c_{s,p}^2}+M_{deg}\frac{1-f\left(\bm{\phi}\right)}{M_{cst}+M_{deg}}\mathbf{c}_i\cdot\eta_p\nabla\bar{\mu}_p\right],
\end{equation}
where $\omega_{i,p}$ is the weight coefficient, $\eta_p$ is an adjustable parameter, $c_{s,p}^2=d_0^pc^2$ is the lattice sound speed with $c=\Delta x/\Delta t$, $d_0^p$ is a scale factor and $\Delta x$ is the lattice spacing. The CH equation (\ref{eq-CH}) can be correctly recovered from above LB model through the direct Taylor (or Chapman-Enskog) expansion with $\left(M_{cst}+M_{deg}\right)/\gamma_p=\left(1/s_1^p-0.5\right)\eta_pc_{s,p}^2\Delta t$, where $s_1^p$ is the relaxation parameter related to an eigenvalue of the collision matrix $(\Lambda_{ij}^p)$.  

We now consider the LB model for NS equations, and compared to our previous works \cite{Zhan2022PRE,Liu2022MMS}, an adjustable scale factor is introduced to make the present model more flexible. The evolution equation of the LB model reads 
\begin{equation}
	g_i\left(\mathbf{x}+\mathbf{c}_i\Delta t,t+\Delta t\right)=g_i\left(\mathbf{x},t\right)-\Lambda_{ij}\left[g_j\left(\mathbf{x},t\right)-g_j^{eq}\left(\mathbf{x},t\right)\right]+\Delta t\left(\delta_{ij}-\frac{\Lambda_{ij}}{2}\right)F_j\left(\mathbf{x},t\right),
\end{equation}
where $g_i\left(\mathbf{x},t\right)$ is the distribution function for fluid field, and the corresponding equilibrium distribution function $g_i^{eq}\left(\mathbf{x},t\right)$ is given by
\begin{equation}
	g_i^{eq}=\lambda_i+\omega_i\frac{\mathbf{c}_i\cdot\rho\mathbf{u}}{c_s^2}+\omega_i\left[\frac{\left(c_{i\alpha}c_{i\alpha}-c_s^2\right)\left(\rho u_{\alpha}u_{\alpha}-S_{\alpha}u_{\alpha}\right)}{c^2c_s^2-c_s^4}+\frac{c_{i\alpha}c_{i\bar{\alpha}}}{c_s^4}\left(\rho u_{\alpha}u_{\bar{\alpha}}-\frac{S_{\alpha}u_{\bar{\alpha}}+u_{\alpha}S_{\bar{\alpha}}}{2}\right)\right],
\end{equation}
with $\lambda_0=\left(\omega_0-1\right)P/c_s^2+\rho_0$, and $\lambda_i=\omega_iP/c_s^2\,(i\neq 0)$. $\omega_i$ and $c_s$ are also the weight coefficient and lattice sound speed, which can be different from those of LB model for CH equation. Here the two terms in the square bracket correspond to the diagonal and non-diagonal parts of the momentum flux, respectively. $\alpha=1,2,\cdots,d$, $\alpha<\bar{\alpha}\leq d$ with $d$ being the dimensionality. The distribution function of force term is expressed as
\begin{equation}
	F_i=\omega_i\left[\mathbf{u}\cdot\nabla\rho+\frac{\mathbf{c}_i\cdot\left(\mathbf{F}+\phi_0\rho\mathbf{f}\right)}{c_s^2}+\frac{\left(c_{i\alpha}c_{i\alpha}-c_s^2\right)M_{\alpha\alpha}^{2F}}{c^2c_s^2-c_s^4}+\frac{c_{i\alpha}c_{i\bar{\alpha}}M_{\alpha\bar{\alpha}}^{2F}}{c_s^4}\right],
\end{equation}
where $\mathbf{F}=\mathbf{F}_s+\mathbf{F}_b+\mathbf{F}_c$, $\mathbf{F}_c=\nabla\cdot\left(\mathbf{S}\mathbf{u}-\mathbf{u}\mathbf{S}\right)/2$ is a corrected force term to exactly recover the consistent momentum equation (\ref{eq-NS2}), and the second-order tensor $\mathbf{M}^{2F}$ is designed as
\begin{equation}
	\mathbf{M}^{2F}=\partial_t\left(\rho\mathbf{u}\mathbf{u}-\frac{\mathbf{S}\mathbf{u}+\mathbf{u}\mathbf{S}}{2}\right)+c_s^2\left(\mathbf{u}\nabla\rho+\nabla\rho\mathbf{u}\right)+\left(c^2-3c_s^2\right)\left(\mathbf{u}\cdot\nabla\rho\right)\mathbf{I}.
\end{equation} 

Through some asymptotic analysis methods, the NS equations (\ref{eq-NS}) can be recovered with the following relation \cite{Chai2023PRE},
\begin{equation}
	\frac{\mu}{\rho\left(1-\phi_0\right)}=\left(\frac{1}{s_{2a}}-\frac{1}{2}\right)\frac{c^2-c_s^2}{2}\Delta t=\left(\frac{1}{s_{2b}}-\frac{1}{2}\right)c_s^2\Delta t,
\end{equation} 
where $s_{2a}$ and $s_{2b}$ are two relaxation parameters and will be stated below. 

Finally, to update the evolution of the LB method, the macroscopic phase-field variable, velocity and pressure are computed by
\begin{equation}
	\phi_p=\sum_if_i,
\end{equation}
\begin{equation}
	\rho\mathbf{u}^*=\sum_i\mathbf{c}_ig_i+\frac{\Delta t}{2}\mathbf{F},\quad \mathbf{u}=\mathbf{u}^*+\frac{\Delta t}{2}\phi_0\mathbf{f},
\end{equation}
\begin{equation}
	p=\frac{c_s^2}{1-\omega_0}\left[\sum_{i\neq0}g_i+\left(\frac{1}{2}+H\right)\Delta t\mathbf{u}\cdot\nabla\rho-\omega_0\frac{\left(\rho\mathbf{u}-\mathbf{S}\right)\cdot\mathbf{u}}{c^2-c_s^2}+\frac{K}{c^2}\frac{\Delta t}{2}\partial_t\left(\left(\rho\mathbf{u}-\mathbf{S}\right)\cdot\mathbf{u}\right)\right],
\end{equation}
where $\mathbf{u}^*$ is the velocity without considering the fluid-solid interaction, $\mathbf{f}$ can be discretized by $\mathbf{f}=\left(\mathbf{u}_w-\mathbf{u}^*\right)/\Delta t$ with $\mathbf{u}_w$ being the velocity of solid point. $K=1-d_0$ and $H=K\left[\left(1-d_0\right)/s_0-\left(2-4d_0\right)/s_{2a}+\left(1-3d_0\right)/2\right]$ for two-dimensional (2D) problem, while $K=1-2d_0$ and $H=K\left[\left(1-d_0\right)/s_0-\left(3-7d_0\right)/s_{2a}+\left(1-3d_0\right)\right]$ for three-dimensional (3D) case, $d_0$ is the adjustable scale factor and $s_0$ is the relaxation parameter corresponding to the zero-order moment of the distribution function. $\partial_t\left(\rho\mathbf{uu}\right)$ is usually approximated by $\mathbf{u}\mathbf{F}+\mathbf{F}\mathbf{u}$, other temporal derivatives are discretized by the explicit Euler scheme, and the gradient and Laplacian operators are computed by the second-order isotropic central schemes, as shown in the previous work \cite{Zhan2022PRE}. Here it is worth noting that when $d_0=1/3$, the present model would reduce to the usual LB model for the two-phase problems.

\section{Numerical simulations and discussion}\label{Numerical}
In the framework of LB method, there are some LB models based on different collision matrices, such as the classical MRT-LB model \cite{Humieres1992,Lallemand2000PRE,Coveney2002}, the cascaded or central-moment LB model \cite{Geier2006PRE,Premnath2012CiCP}, the Hermite moment LB model \cite{Coreixas2017PRE}, the central-Hermite moment LB model \cite{Coreixas2019PRE,Mattila2017PF} and so on. However, under the unified framework of modeling of the MRT-LB model in Refs. \cite{Chai2020PRE,Chai2023PRE}, different forms of the collision matrices can be converted into each other through a specific lower triangular matrix with the diagonal element of unity. Therefore, when conducting the collision step of LB method in moment space to improve the efficiency of the algorithm, one can use the natural moment space, and multiply a specific lower triangular matrix on the relaxation matrix to apply the corresponding collision matrix. For example, the collision matrix of the Hermite moment LB model $\bm{\Lambda}=\mathbf{H}^{-1}\mathbf{KH}$ ($\mathbf{H}$ is the Hermite transformation matrix and $\mathbf{K}$ is the diagonal relaxation matrix) can be written as
\begin{equation}
	\bm{\Lambda}=\left(\mathbf{NM}\right)^{-1}\mathbf{KNM}=\mathbf{M}_0^{-1}\left(\mathbf{C}_d^{-1}\mathbf{N}^{-1}\mathbf{KN}\mathbf{C}_d\right)\mathbf{M}_0=\mathbf{M}_0^{-1}\mathbf{S}\mathbf{M}_0,
\end{equation}
where $\mathbf{M}=\mathbf{C}_d\mathbf{M}_0$ is the natural transformation matrix with $\mathbf{C}_d$ being a diagonal matrix formed by the powers of the lattice speed $c$, $\mathbf{N}$ is the lower triangular matrix associated with $\mathbf{H}$ and $\mathbf{M}$, and $\mathbf{S}$ is a new lower triangular relaxation matrix. 

In this work, we will apply the above Hermite moment LB model to perform some 2D and 3D simulations, and the specific forms of weight coefficients $\omega_{i,p}$ and $\omega_i$, matrices $\mathbf{K}$, $\mathbf{C}_d$, $\mathbf{M}_0$, and $\mathbf{N}$ in D2Q9 and D3Q15 lattice structures are given in \ref{Matrices}. In the numerical simulations, the lattice units are used, and are listed in Table \ref{tab-units}. The half-way bounce-back scheme \cite{Ladd1994JFM} is applied for the no-flux scalar and no-slip velocity boundary conditions, and the non-equilibrium extrapolation scheme \cite{Guo2002CP} for the velocity or pressure boundary conditions at inlet and/or outlet. 
Additionally, we apply the degenerate mobility by setting $f\left(\bm{\phi}\right)=\prod_{p=0}^2\left(1-\phi_p\right)$ for \emph{3f} model and $f\left(\bm{\phi}\right)=\phi_1\left(1-\phi_1\right)\left(1-\phi_0\right)$ for \emph{2f1s} model. In the fluid region with $\phi_0=0$, the present $f\left(\bm{\phi}\right)$ in \emph{2f1s} model can reduce to the classical degenerate mobility, which indicates that the boundedness of the solution of CH equation for two-phase flows can be preserved \cite{Elliott1996JMA}.  

\begin{table}
	\centering
	\caption{Units in lattice Boltzmann method}
	\begin{tabular}{ccccccc}
		\toprule
		Variable && Mass && Length && Time \\
		\midrule
		Unit && \si{mu} && \si{lu} && \si{ts}\\
		\bottomrule
	\end{tabular}
	\label{tab-units}
\end{table}

\subsection{The results of two models for a simple case}
As mentioned earlier, there are some differences between the \emph{3f} model and \emph{2f1s} model, i.e., the values of solid/liquid surface tension coefficients, the number of the CH equation, the expressions of the degenerate mobility and the effective viscosity, and the first two of them are the key limitations from \emph{3f} model to \emph{2f1s} model. We first show the numerical results of these two models through performing a simple test in the 2D domain of $[-L_0/2,L_0/2]\times[-L_0/2,L_0/2]$ with $L_0=300$ \si{[lu]}. The periodic boundary condition is applied at the left and right boundaries, while the no-flux boundary condition is imposed on the upper and lower walls. A droplet or a neutral particle with the radius $R_0=50$ \si{[lu]} is placed at the interface of a binary fluid layer, and the phase-field variables are initialized by
\begin{equation}
	\begin{aligned}
		\phi_0\left(x,y\right)&=\frac{1}{2}+\frac{1}{2}\tanh\frac{R_0-\sqrt{x^2+y^2}}{D/2},\\
		\phi_1\left(x,y\right)&=\left[1-\phi_0\left(x,y\right)\right]\left(\frac{1}{2}-\frac{1}{2}\tanh\frac{2y}{D}\right).
	\end{aligned}
\end{equation}
When the gravity is not considered, the droplet will form a lens at the equilibrium state dominated by the surface tension, while the neutral particle will maintain the initial state. 

\begin{figure}
	\centering
	\subfigure[Droplet based on \emph{3f} model]{
		\begin{minipage}{0.48\linewidth}
			\centering
			\includegraphics[width=1.8in]{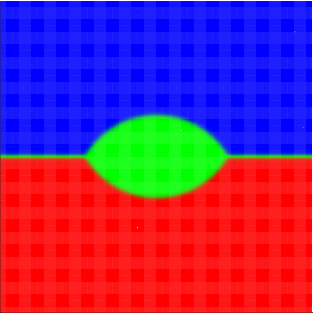}
	\end{minipage}}
	\subfigure[Neutral particle based on \emph{2f1s} model]{
		\begin{minipage}{0.48\linewidth}
			\centering
			\includegraphics[width=1.8in]{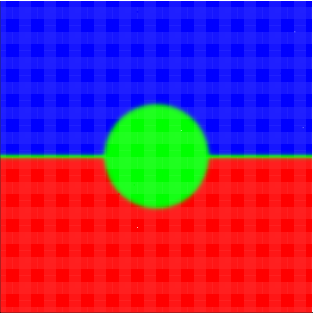}
	\end{minipage}}
	\caption{The equilibrium shapes of the droplet and the neutral particle.}
	\label{fig-com2model}
\end{figure}

In our simulations, we set some physical parameters as $M_{cst}=0$, $M_{deg}=0.1$, $\rho_0:\rho_1:\rho_2=5:10:1$ \si{[mu/lu^3]}, $\mu_0:\mu_1:\mu_2=0.05:0.1:0.01$ \si{[mu/(lu.ts)]}, $D=5$ \si{[lu]} and $\sigma_{12}=0.001$ \si{[mu/ts^2]}. $\sigma_{01}=\sigma_{02}=\sigma_{12}$ is adopted for the \emph{3f} model, while Eq. (\ref{eq-kk}) is used for \emph{2f1s} model. The final shapes of droplet and neutral particle are plotted in Fig. \ref{fig-com2model}. From this figure, one can observe that these two models show a good performance for the multiphase problems.

\begin{figure}
	\centering
	\includegraphics[width=4.5in]{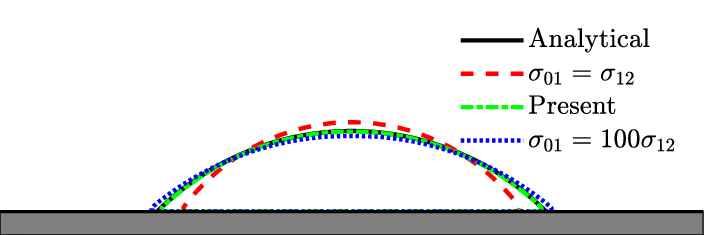}
	\caption{The equilibrium shapes of the droplet on an ideal wall with different values of $\sigma_{01}$.}
	\label{fig-plateK}
\end{figure}
\begin{figure}
	\centering
	\subfigure[$\theta=60^\circ$]{
		\begin{minipage}{0.48\linewidth}
			\centering
			\includegraphics[width=3.0in]{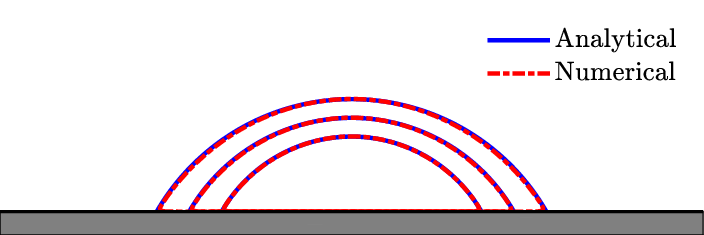}
	\end{minipage}}
	\subfigure[$\theta=120^\circ$]{
		\begin{minipage}{0.48\linewidth}
			\centering
			\includegraphics[width=3.0in]{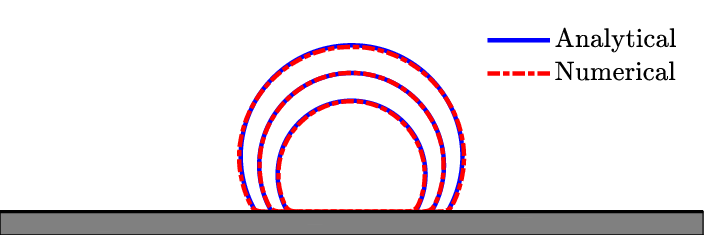}
	\end{minipage}}
	\caption{The equilibrium shapes of the droplet on an ideal wall with different initial radii ($R_0=40,50,60$ \si{[lu]}).}
	\label{fig-plateR}
\end{figure}
\begin{figure}
	\centering
	\subfigure[$\theta=30^\circ$]{
		\begin{minipage}{0.48\linewidth}
			\centering
			\includegraphics[width=3.0in]{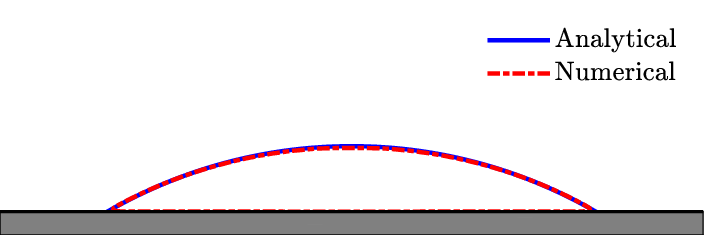}
	\end{minipage}}
	\subfigure[$\theta=45^\circ$]{
		\begin{minipage}{0.48\linewidth}
			\centering
			\includegraphics[width=3.0in]{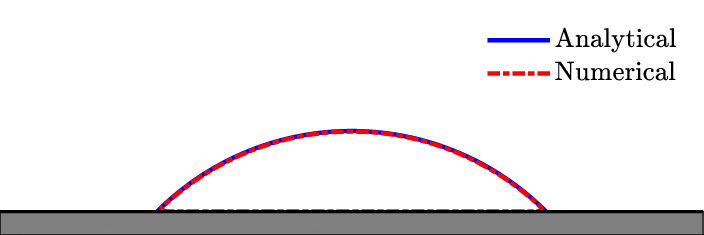}
	\end{minipage}}
	
	\subfigure[$\theta=135^\circ$]{
		\begin{minipage}{0.48\linewidth}
			\centering
			\includegraphics[width=3.0in]{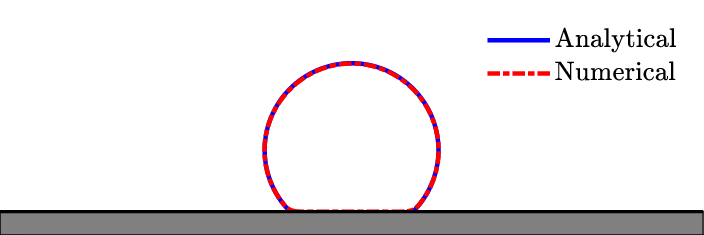}
	\end{minipage}}
	\subfigure[$\theta=150^\circ$]{
		\begin{minipage}{0.48\linewidth}
			\centering
			\includegraphics[width=3.0in]{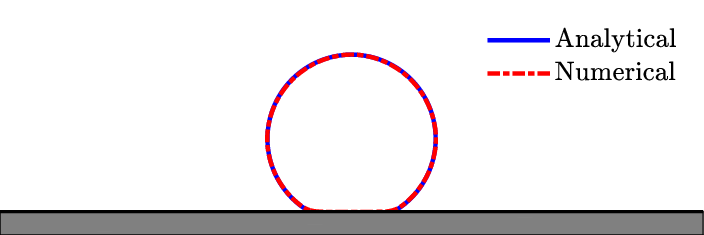}
	\end{minipage}}
	\caption{The equilibrium shapes of the droplet on an ideal wall at different contact angles.}
	\label{fig-plate}
\end{figure}
\begin{table}
	\centering
	\caption{The normalized droplet heights at different contact angles.}
	\begin{tabular}{cccccccccc}
		\toprule
		Contact angle & $30^\circ$ & $45^\circ$ & $60^\circ$ & $75^\circ$ & $90^\circ$ & $105^\circ$ & $120^\circ$ & $135^\circ$ & $150^\circ$\\
		\midrule
		Analytical & 0.5427 & 0.6822 & 0.8008 & 0.9013 & 1.0000 & 1.0985 & 1.1830 & 1.2628 & 1.3393 \\
		Numerical & 0.5579 & 0.6871 & 0.7996 & 0.9027 & 1.0000 & 1.0933 & 1.1825 & 1.2660 & 1.3389 \\
		Relative error & 2.72\% & 0.72\% & 0.15\% & 0.16\% & 0 & 0.48\% & 0.04\% & 0.25\% & 0.03\% \\ 
		\bottomrule
	\end{tabular}
	\label{tab-plate}
\end{table}
\subsection{A droplet spreading on an ideal wall}\label{plate}
To test the accuracy of the present \emph{2f1s} model, here we consider a droplet spreading on an idea wall, and the solid plate with a finite width $H_0=10$ \si{[lu]} is absorbed into the computational domain $[-L_0/2,L_0/2]\times[-H_0,9H_0]$ with $L_0=300$ \si{[lu]}. A semicircular droplet with the radius $R_0$ is placed on the center of the wall, the periodic boundary condition is applied in $x$-direction and no-flux boundary condition is imposed on the top and bottom boundaries. Initially, the phase-field variables are given by 
\begin{equation}
	\begin{aligned}
		\phi_0\left(x,y\right)&=\frac{1}{2}-\frac{1}{2}\tanh\frac{2y}{D},\\
		\phi_1\left(x,y\right)&=\left[1-\phi_0\left(x,y\right)\right]\left(\frac{1}{2}+\frac{1}{2}\tanh\frac{R_0-\sqrt{x^2+y^2}}{D/2}\right).
	\end{aligned}
\end{equation}
In the simulations, some physical parameters are fixed as $\rho_0:\rho_1:\rho_2=1$ \si{[mu/lu^3]}, $\mu_0=\mu_1=\mu_2=0.01$ \si{[mu/(lu.ts)]} and $\sigma_{12}=0.001$ \si{[mu/ts^2]}. We first give a comparison of the results under different values of $\sigma_{01}$ in Fig. \ref{fig-plateK}. From this figure, one can observe that the spreading of the droplet is not enough with a small $k=\sigma_{01}/\sigma_{12}$, while a large value of $k$ leads to an inaccurate result, although the mechanical balance condition (\ref{eq-triple}) is preserved better. We believe that there should be an optimal value of $k$ to give the best result, and the proposed equation (\ref{eq-kk}) seems to present a good performance, as shown in Fig. \ref{fig-plateK}. We note that in this case, the curvature of the solid plate is zero, and the parameter $k$ of Eq. (\ref{eq-kk}) is only related to the contact angle. To further test the accuracy of Eq. (\ref{eq-kk}), the droplet spreading on the solid plate with different initial radii are considered. As seen from Fig. \ref{fig-plateR}, the present numerical results agree well with the analytical solutions. We also plot the results at a large range of the contact angles in Fig. \ref{fig-plate} where the initial radius of the droplet is 50 \si{[lu]}, and find that the finial shapes of the droplet are exactly determined by the given contact angles. To give a quantitative comparison, we also measure the height of the droplet from the equilibrium shape in Table \ref{tab-plate}, and it is found that the numerical data are in good agreement with the following analytical solution,
\begin{equation}
	H=R-R\cos\theta,\quad R=R_0\sqrt{\frac{\pi/2}{\theta-\sin\theta\cos\theta}}.
\end{equation}
Here it should be noted that the droplet spreading length on the solid surface is not measured because the diffuse-interface method for multiphase system usually loses a certain accuracy at triple points. 

\begin{figure}
	\centering
	\subfigure[$\theta=45^\circ$]{
		\begin{minipage}{0.48\linewidth}
			\centering
			\includegraphics[width=2.0in]{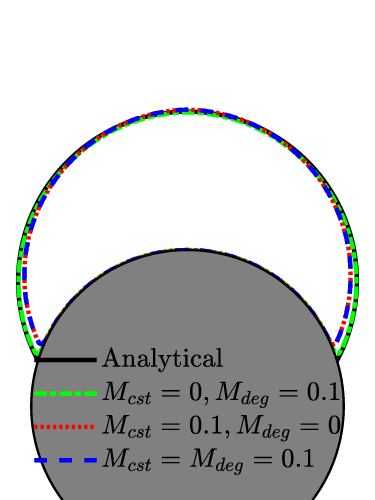}
	\end{minipage}}
	\subfigure[$\theta=135^\circ$]{
		\begin{minipage}{0.48\linewidth}
			\centering
			\includegraphics[width=2.0in]{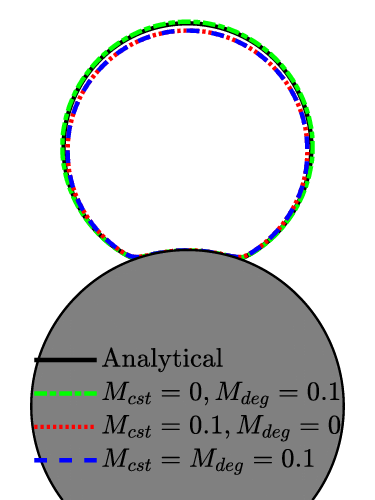}
	\end{minipage}}
	\caption{The equilibrium shapes of the droplet on a cylinder surface with different forms of mobility.}
	\label{fig-cylinderM}
\end{figure}
\begin{figure}
	\centering
	\subfigure[$R_0=75$ \si{[lu]}, $R_s=50$ \si{[lu]}]{
		\begin{minipage}{0.3\linewidth}
			\centering
			\includegraphics[width=1.5in]{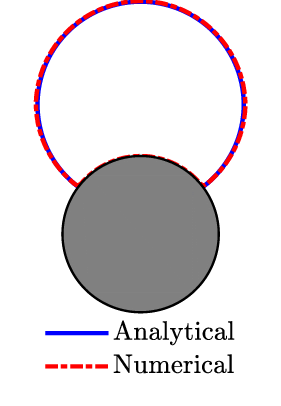}
	\end{minipage}}
	\subfigure[$R_0=50$ \si{[lu]}, $R_s=50$ \si{[lu]}]{
		\begin{minipage}{0.3\linewidth}
			\centering
			\includegraphics[width=1.5in]{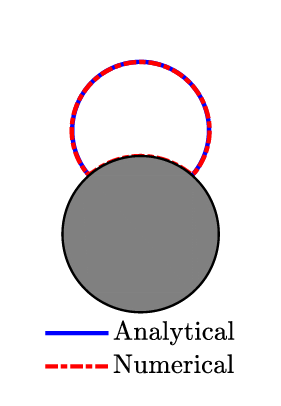}
	\end{minipage}}
	\subfigure[$R_0=50$ \si{[lu]}, $R_s=75$ \si{[lu]}]{
		\begin{minipage}{0.3\linewidth}
			\centering
			\includegraphics[width=1.5in]{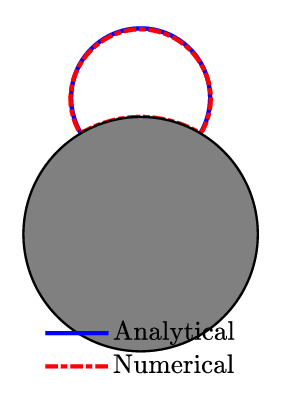}
	\end{minipage}}
	\caption{The equilibrium shapes of the droplet on a cylinder surface with different radius ratios and $\theta=90^\circ$.}
	\label{fig-cylinderR}
\end{figure}
\begin{figure}
	\centering
	\subfigure[$\theta=30^\circ$]{
		\begin{minipage}{0.24\linewidth}
			\centering
			\includegraphics[width=1.5in]{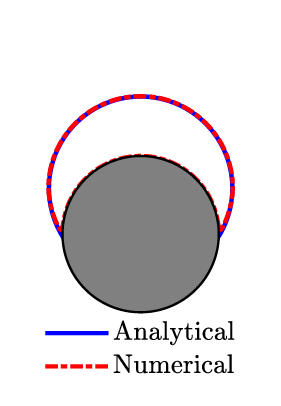}
	\end{minipage}}
	\subfigure[$\theta=60^\circ$]{
		\begin{minipage}{0.24\linewidth}
			\centering
			\includegraphics[width=1.5in]{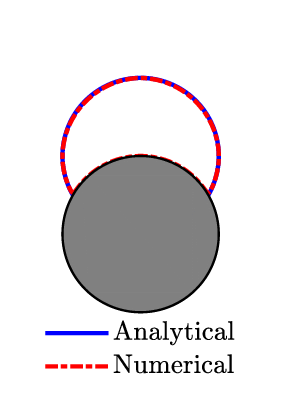}
	\end{minipage}}	
	\subfigure[$\theta=120^\circ$]{
		\begin{minipage}{0.24\linewidth}
			\centering
			\includegraphics[width=1.5in]{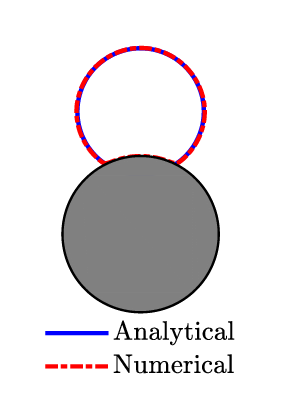}
	\end{minipage}}
	\subfigure[$\theta=150^\circ$]{
		\begin{minipage}{0.24\linewidth}
			\centering
			\includegraphics[width=1.5in]{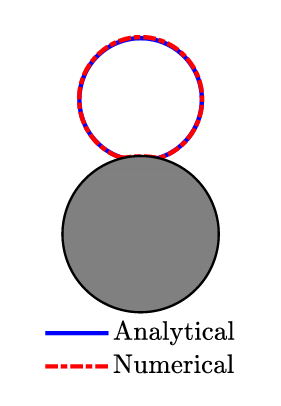}
	\end{minipage}}
	\caption{The equilibrium shapes of the droplet on a cylinder surface at different contact angles.}
	\label{fig-cylinder}
\end{figure}
\begin{figure}
	\centering
	\includegraphics[width=3.5in]{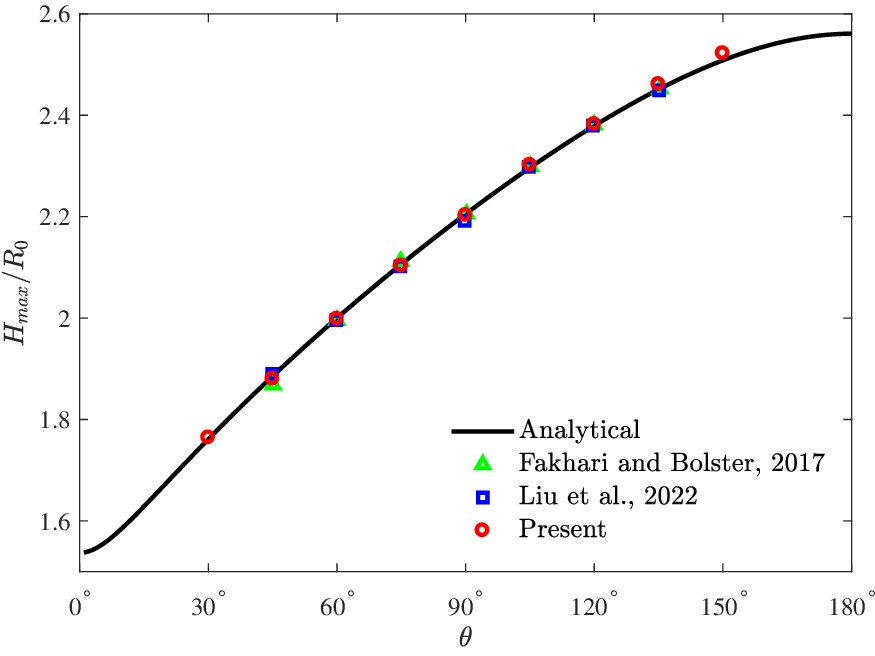}
	\caption{The normalized droplet heights at different contact angles.}
	\label{fig-cylinderH}
\end{figure}
\subsection{A droplet spreading on a cylinder surface}\label{cylinder}
In this part, we consider a droplet spreading on a cylinder surface in the square domain $[-L_0/2,L_0/2]\times[-L_0/4,3L_0/4]$ with $L_0=300$ \si{[lu]}, and the other physical parameters are the same as those in the previous section. In the computational domain, a solid cylinder with the radius $R_s$ is located at the position $\left(0,0\right)$, and a droplet with radius $R_0$ is placed on the cylinder surface with the following initial phase-field variables,
\begin{equation}
	\begin{aligned}
		\phi_0\left(x,y\right)&=\frac{1}{2}+\frac{1}{2}\tanh\frac{R_s-\sqrt{x^2+y^2}}{D/2},\\
		\phi_1\left(x,y\right)&=\left[1-\phi_0\left(x,y\right)\right]\left[\frac{1}{2}+\frac{1}{2}\tanh\frac{R_0-\sqrt{x^2+\left(y-R_s\right)^2}}{D/2}\right].
	\end{aligned}
\end{equation}

For this problem, the curvature of the cylinder is $1/R_s$, and thus we set the dimensionless curvature to be $\kappa=R_0/R_s$. 
Usually, a constant mobility in the CH equation for two-phase flows cannot preserve the boundedness of the phase-field variable and the phase mass conservation. However, a degenerate mobility can overcome these defects \cite{Elliott1996JMA}. Here we present a comparison between different forms of mobility through considering the droplet spreading on a cylinder surface, and show the results in Fig. \ref{fig-cylinderM}. From this figure, one can find that the constant part of mobility causes the obvious shrinkage of the droplet, and there is a mass leakage between the two fluid phases, although the total mass is conserved. For this reason, the degenerate mobility with $M_{cst}=0$ is used in the following simulations. On the other hand, we also test Eq. (\ref{eq-kk}) by considering the effect of dimensionless curvature $\kappa$. As shown in Fig. \ref{fig-cylinderR}, the numerical results also match well with the analytical solutions. Additionally, we also present the equilibrium shapes of the droplet at a large range of contact angles in Fig. \ref{fig-cylinder}, and find that the droplets experience the greater deformations, compared to those on the plate walls in Section \ref{plate}. To conduct a quantitative comparison, we further consider the maximum height of the equilibrium droplet $H_{max}$, and it can be obtained by the initial shape under the condition of equal area,
\begin{equation}
	H_{max}=r+k_s,\quad \pi r^2-r^2\cos^{-1}\left(\frac{k_s^2+r^2-R_s^2}{2k_sr}\right)-R_s^2\cos^{-1}\left(\frac{k_s^2-r^2+R_s^2}{2k_sR_s}\right)+k_s\sqrt{R_s^2-\left(\frac{k_s^2-r^2+R_s^2}{2k_s}\right)}=S_0,
\end{equation}
where $S_0$ is the initial area, $k_s$ is related to the finial radius $r$ and the contact angle $\theta$,
\begin{equation}
	k_s=\sqrt{r^2+R_s^2-2rR_s\cos\theta}.
\end{equation}
Figure \ref{fig-cylinderH} plots the corresponding normalized maximum height of the droplet, and it is found that the present results agree well with the analytical solutions and previous data \cite{Fakhari2017JCP,Liu2022MMS}. 

\begin{figure}
	\centering
	\includegraphics[width=3.5in]{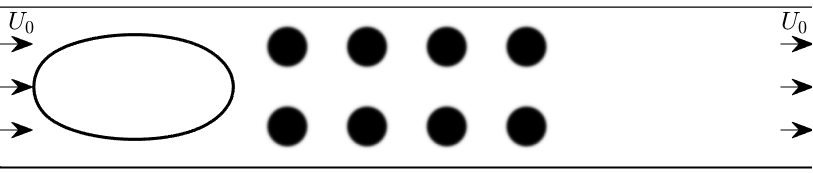}
	\caption{The configuration of a droplet passing through $2\times4$ cylindrical obstacles.}
	\label{fig-obstacles0}
\end{figure}

\subsection{A droplet passing through $2\times4$ cylindrical obstacles}\label{obstacles}
\begin{figure}
	\centering
	\subfigure[]{
		\begin{minipage}{0.48\linewidth}
			\centering
			\includegraphics[width=2.5in]{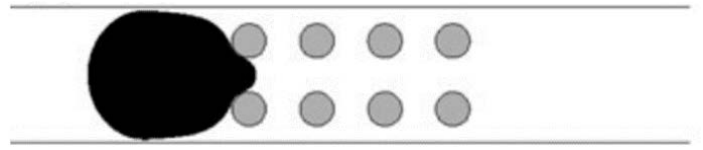}
	\end{minipage}}
	\subfigure[]{
		\begin{minipage}{0.48\linewidth}
			\centering
			\includegraphics[width=2.4in]{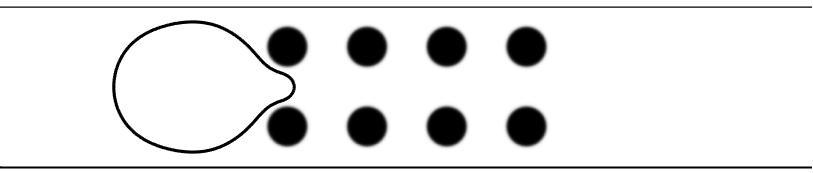}
	\end{minipage}}

	\subfigure[]{
		\begin{minipage}{0.48\linewidth}
			\centering
			\includegraphics[width=2.5in]{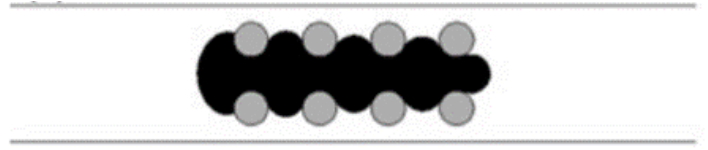}
	\end{minipage}}
	\subfigure[]{
		\begin{minipage}{0.48\linewidth}
			\centering
			\includegraphics[width=2.4in]{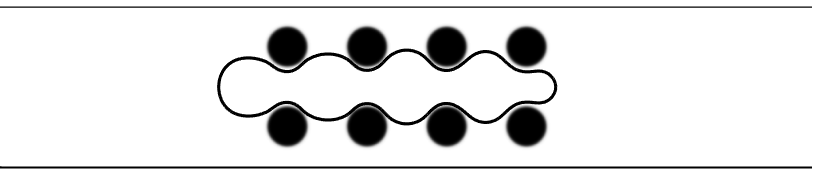}
	\end{minipage}}	
	
	\subfigure[]{
		\begin{minipage}{0.48\linewidth}
			\centering
			\includegraphics[width=2.5in]{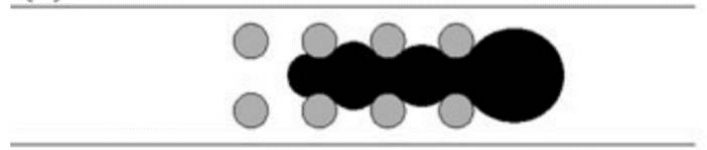}
	\end{minipage}}
	\subfigure[]{
		\begin{minipage}{0.48\linewidth}
			\centering
			\includegraphics[width=2.4in]{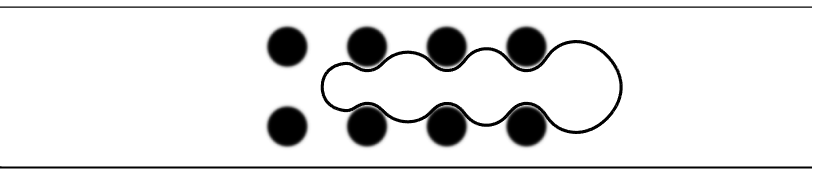}
	\end{minipage}}

	\subfigure[]{
		\begin{minipage}{0.48\linewidth}
			\centering
			\includegraphics[width=2.4in]{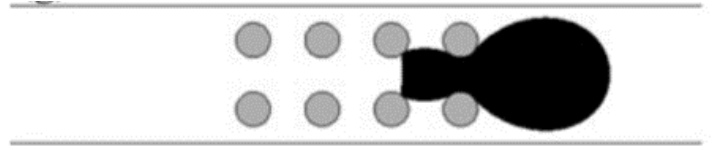}
	\end{minipage}}
	\subfigure[]{
		\begin{minipage}{0.48\linewidth}
			\centering
			\includegraphics[width=2.5in]{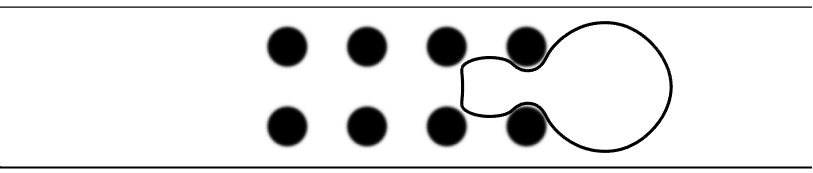}
	\end{minipage}}
	\caption{The snapshots of the droplet passing through $2\times4$ cylindrical obstacles [Results reported in Ref. \cite{Chung2010MN}: (a), (c), (e), (g), present results: (b), (d), (f), (h)].}
	\label{fig-obstacles}
\end{figure}

We now study the problem of a droplet passing through $2\times4$ cylindrical obstacles with a large topological interface change. The cylindrical obstacles with the radius $R_s=25$ \si{[lu]} and center distance $4R_s$ are placed at the central of the channel, the droplet is initialized on the left side of the obstacles, and is driven by the left inlet and right outlet velocity $U_0$ (see the schematic in Fig. \ref{fig-obstacles0}). The physical parameters are set to be $M_{deg}=0.01$, $\rho_0:\rho_1:\rho_2=1$ \si{[mu/lu^3]}, $\mu_0=\mu_1=\mu_2=0.01$ \si{[mu/(lu.ts)]}, $\sigma_{12}=0.001$ \si{[mu/ts^2]}, $\theta=170^\circ$, and Capillary number $Ca=\mu_1U_0/\sigma_{12}=0.01$. We conduct some simulations with the channel length $L=2400$ \si{[lu]} and width $H=200$ \si{[lu]}, and compare present results with the available data \cite{Chung2010MN} in Fig. \ref{fig-obstacles}. From this figure, one can see that present results show the similar dynamic behaviors to those reported in the previous work. Here it should be noted that the initial size of the droplet is not clearly specified in Ref. \cite{Chung2010MN}, and there may be a difference in the droplet size. From this test, one can conclude that the present model has a good capability in the study of two-phase flows in complex geometries, and in the following, it will be adopted to investigate two displacement problems. 

\subsection{Displacement in a Tesla valve}\label{Tesla}
\begin{figure}
	\centering
	\includegraphics[width=2.5in]{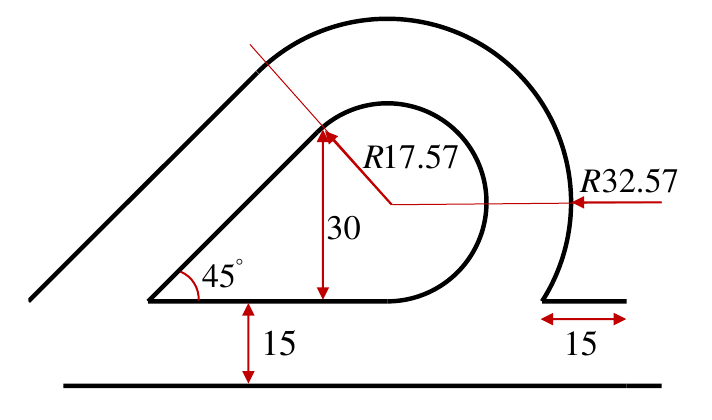}
	\caption{The 2D projected structure of the one stage T45C Tesla valve.}
	\label{fig-Tesla2D}
\end{figure}

\begin{figure}
	\centering
	\centering
	\subfigure[Forward flow]{
		\begin{minipage}{0.48\linewidth}
			\centering
			\includegraphics[width=3.2in]{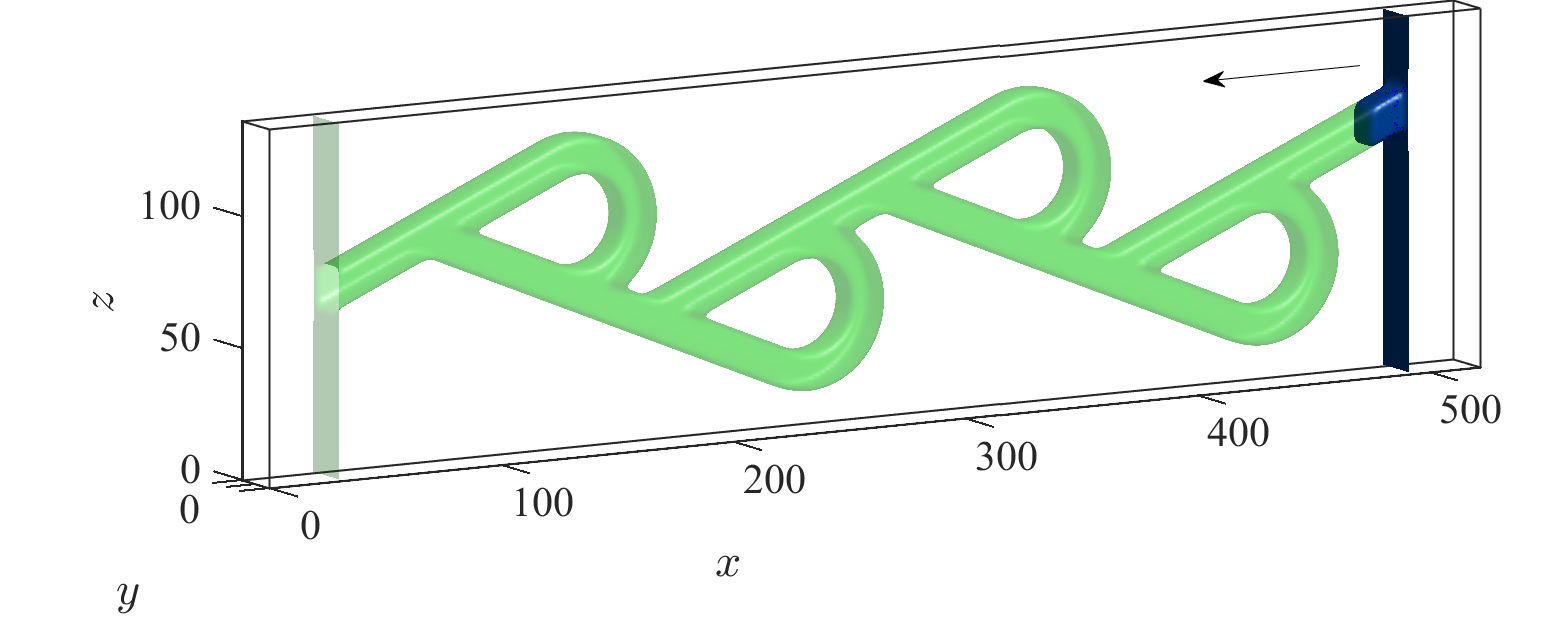}
	\end{minipage}}
	\subfigure[Reverse flow]{
		\begin{minipage}{0.48\linewidth}
			\centering
			\includegraphics[width=3.2in]{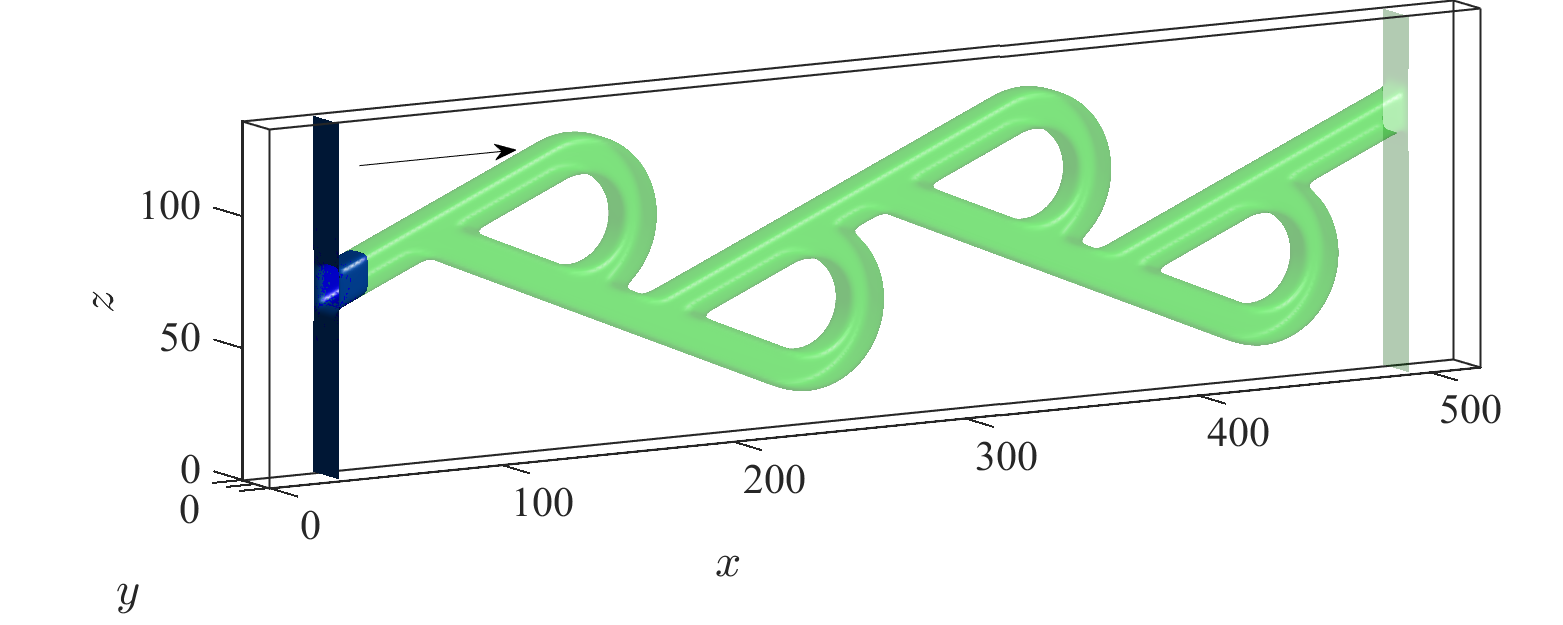}
	\end{minipage}}
	\caption{The 3D structure of the 4 stages T45C Tesla valve and the initialization of driving phase.}
	\label{fig-Tesla3D}
\end{figure}

Tesla valve is a kind of check valve with no-moving-part, invented by Nicolas Tesla in 1920 \cite{Tesla}. This special pipe structure usually presents different flow patterns when the fluid flows in forward and reverse directions, and can be considered as a fluidic diode. Here we consider the displacement of two-phase flow in a 4 stages 3D T45C Tesla valve, and present the 2D projected structure of 1 stage T45C Tesla valve in Fig. \ref{fig-Tesla2D} where the pipe width is $d_s=15$ \si{[lu]} with 1 \si{[lu]}$=4\times10^{-4}$ \si{[m]}, and other structure parameters can be found in Ref. \cite{Bao2022IJHMT}. 
The material properties of the driving fluid (phase 1) and displaced fluid (phase 2) are set to be the same for simplicity, i.e., $\rho_1=\rho_2=1$ \si{[mu/lu^3]}, $\mu_1=\mu_2=0.1$ \si{[mu/(lu.ts)]}, and other parameters are fixed as $D=4$ \si{[lu]}, $\sigma_{12}=0.004$ \si{[mu/ts^2]} and $\theta=60^\circ$. 
In our simulations, the computational domain is $\left(462+2\Delta\right)\times20\times136$ \si{[lu^3]} with $\Delta=30$ being the number of fluid layers added to the two sides of the domain in $x$-direction, as shown in Fig. \ref{fig-Tesla3D}.

\begin{figure}
	\centering
	\subfigure[$t=5\times10^4$ \si{[ts]}]{
		\begin{minipage}{0.48\linewidth}
			\centering
			\includegraphics[width=3.0in]{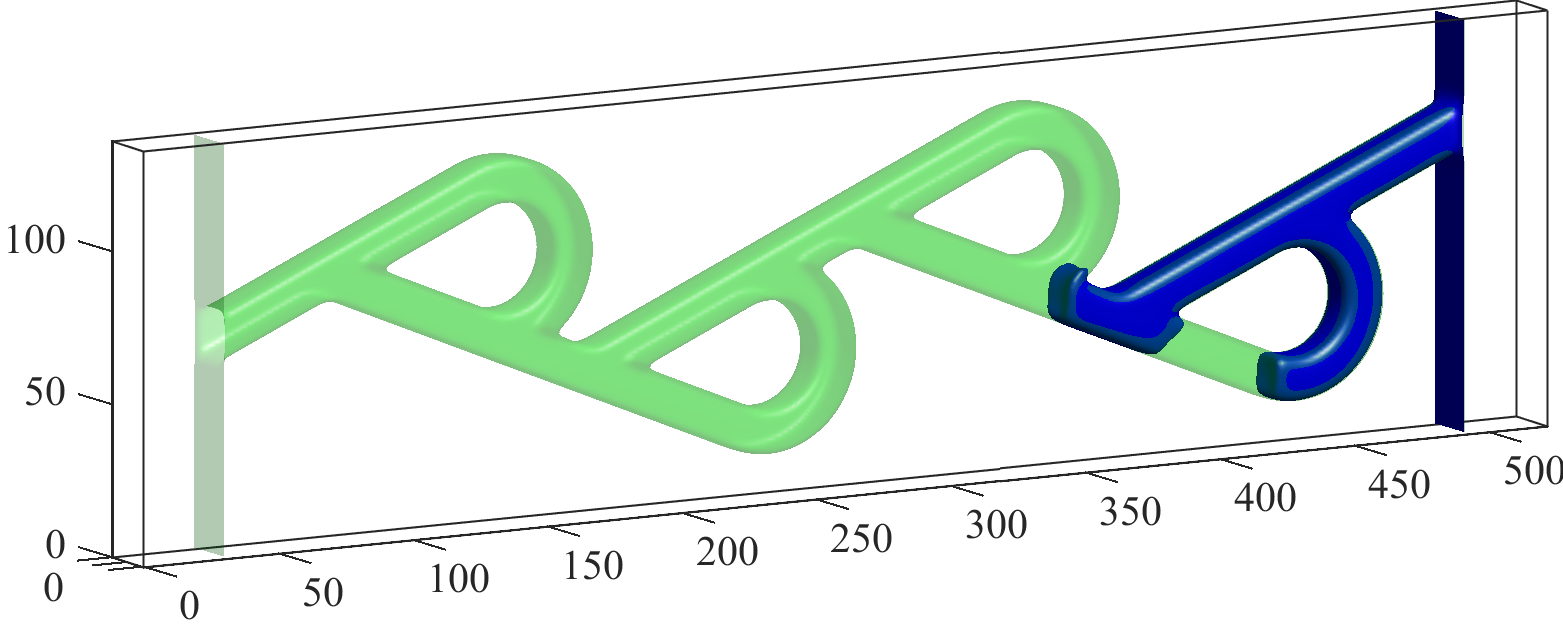}
	\end{minipage}}
	\subfigure[$t=1\times10^5$ \si{[ts]}]{
		\begin{minipage}{0.48\linewidth}
			\centering
			\includegraphics[width=3.0in]{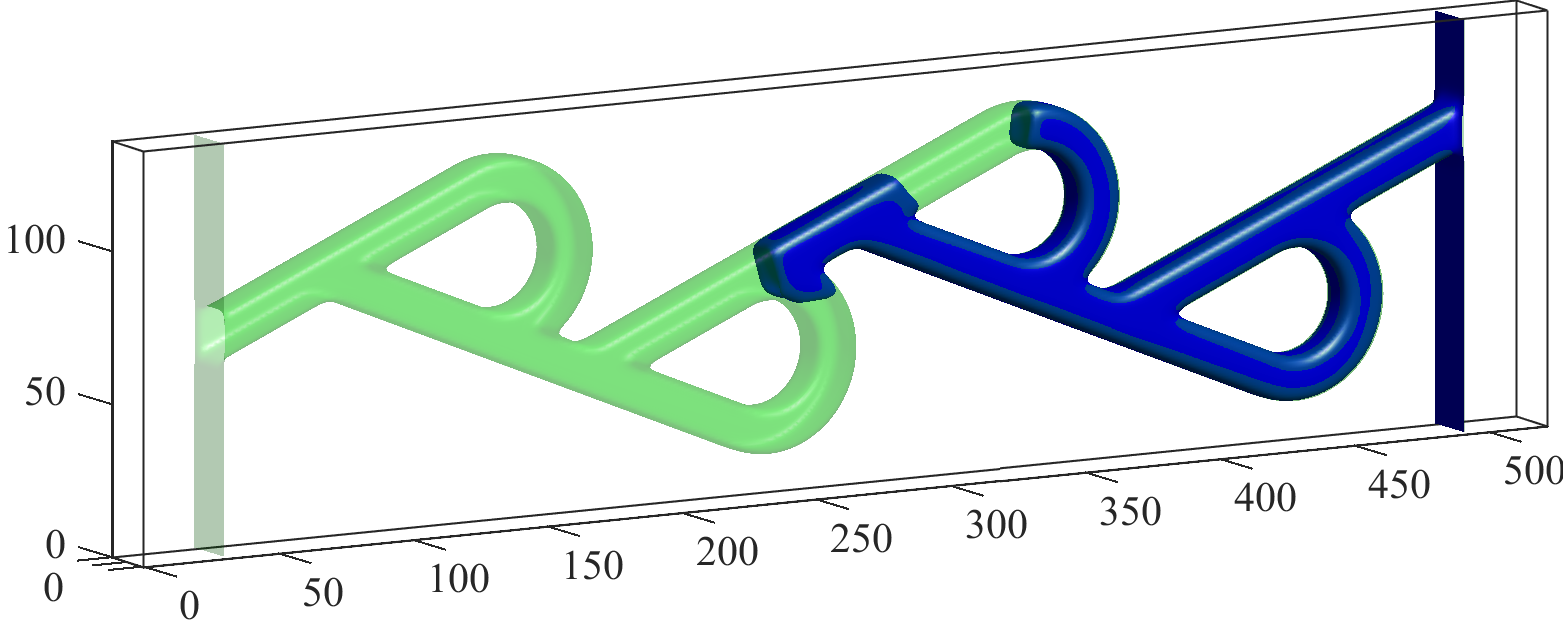}
	\end{minipage}}
	
	\subfigure[$t=1.5\times10^5$ \si{[ts]}]{
		\begin{minipage}{0.48\linewidth}
			\centering
			\includegraphics[width=3.0in]{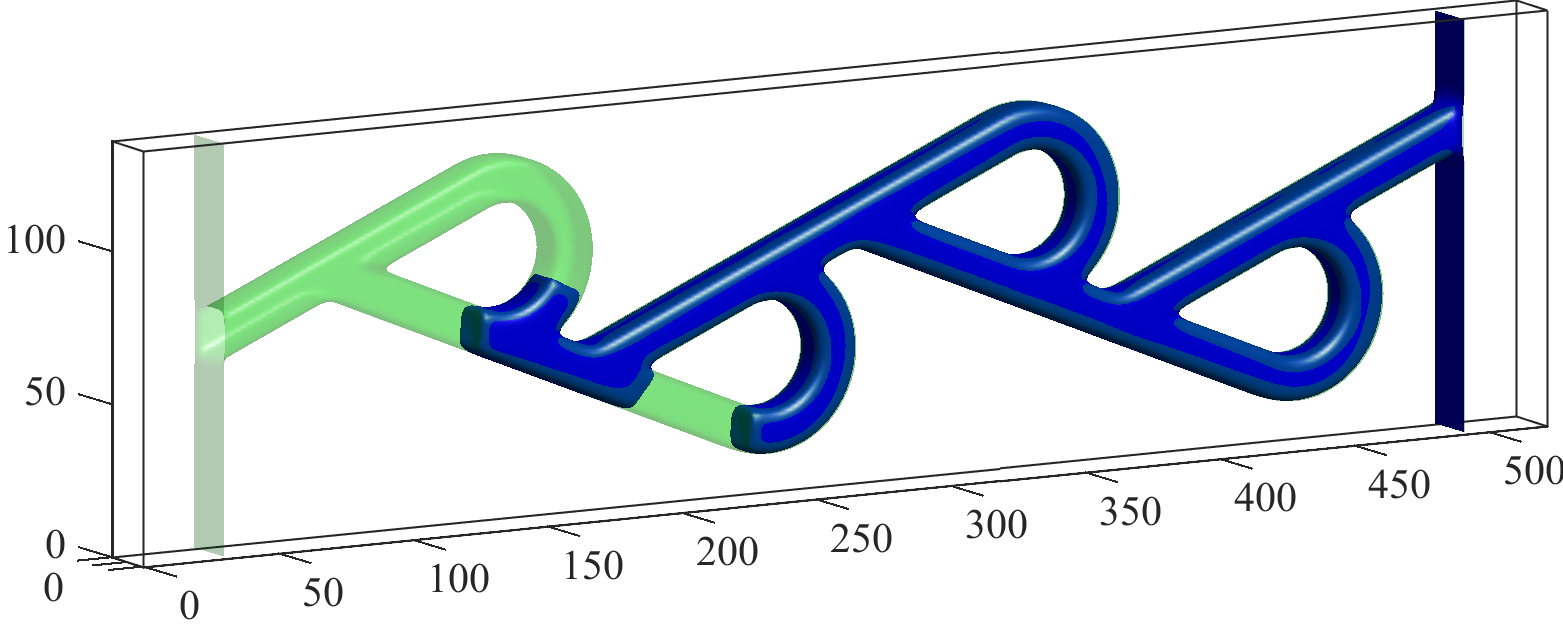}
	\end{minipage}}
	\subfigure[$t=2\times10^5$ \si{[ts]}]{
		\begin{minipage}{0.48\linewidth}
			\centering
			\includegraphics[width=3.0in]{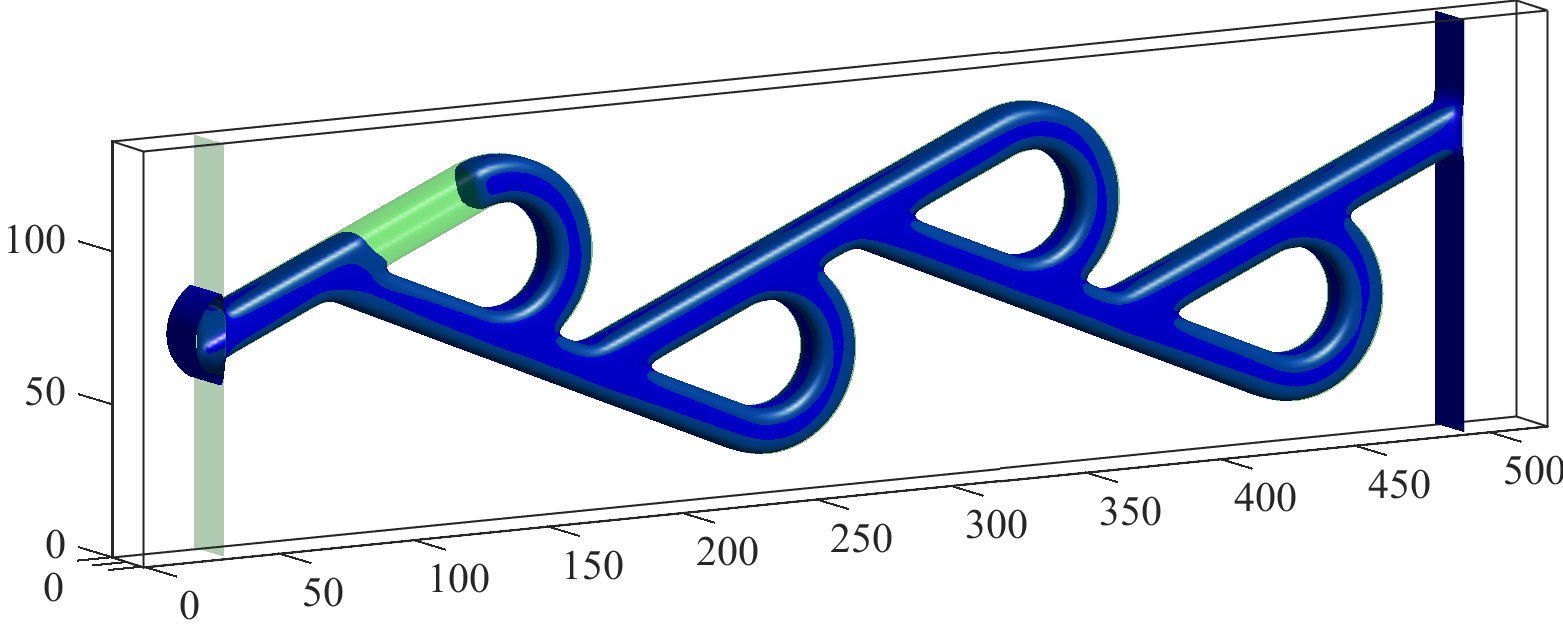}
	\end{minipage}}
	\caption{The snapshots of the forward displacement in the 4 stages T45C Tesla valve.}
	\label{fig-Tesla3DForward}
\end{figure}
\begin{figure}
	\centering
	\subfigure[$t=5\times10^4$ \si{[ts]}]{
		\begin{minipage}{0.48\linewidth}
			\centering
			\includegraphics[width=3.0in]{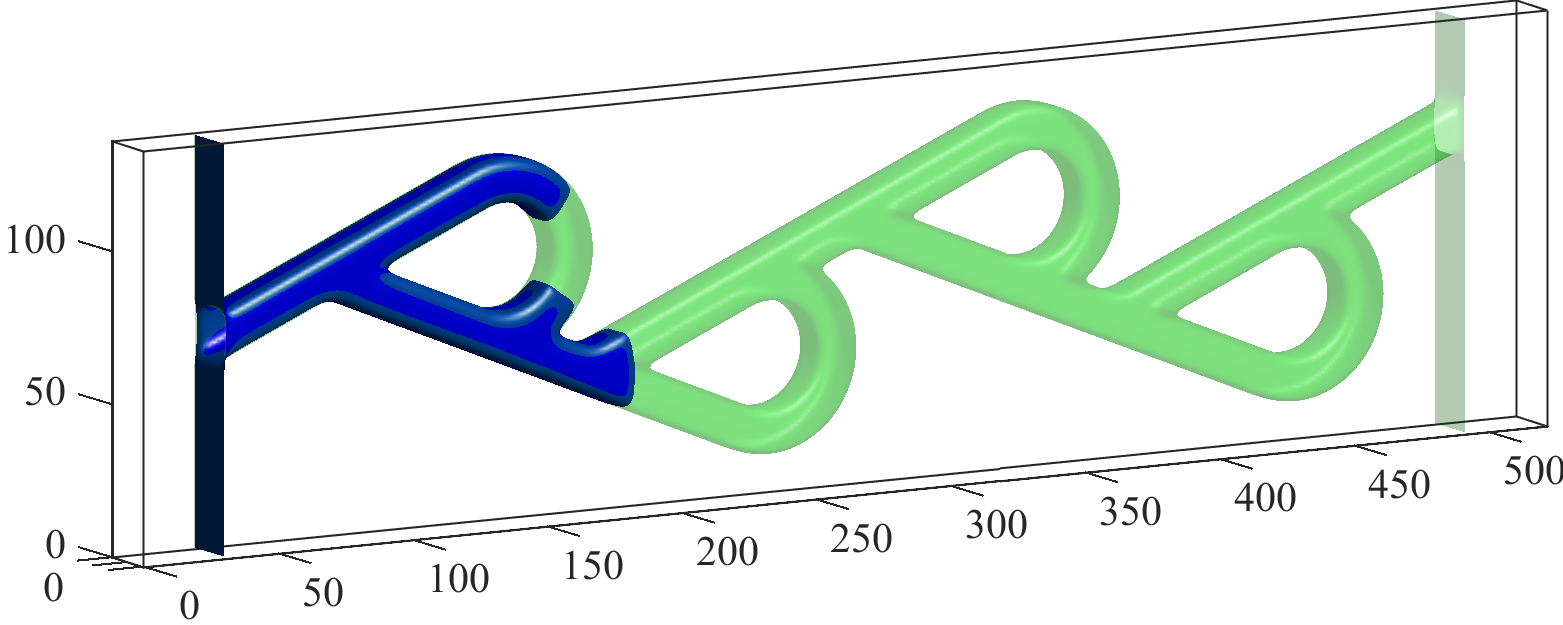}
	\end{minipage}}
	\subfigure[$t=1\times10^5$ \si{[ts]}]{
		\begin{minipage}{0.48\linewidth}
			\centering
			\includegraphics[width=3.0in]{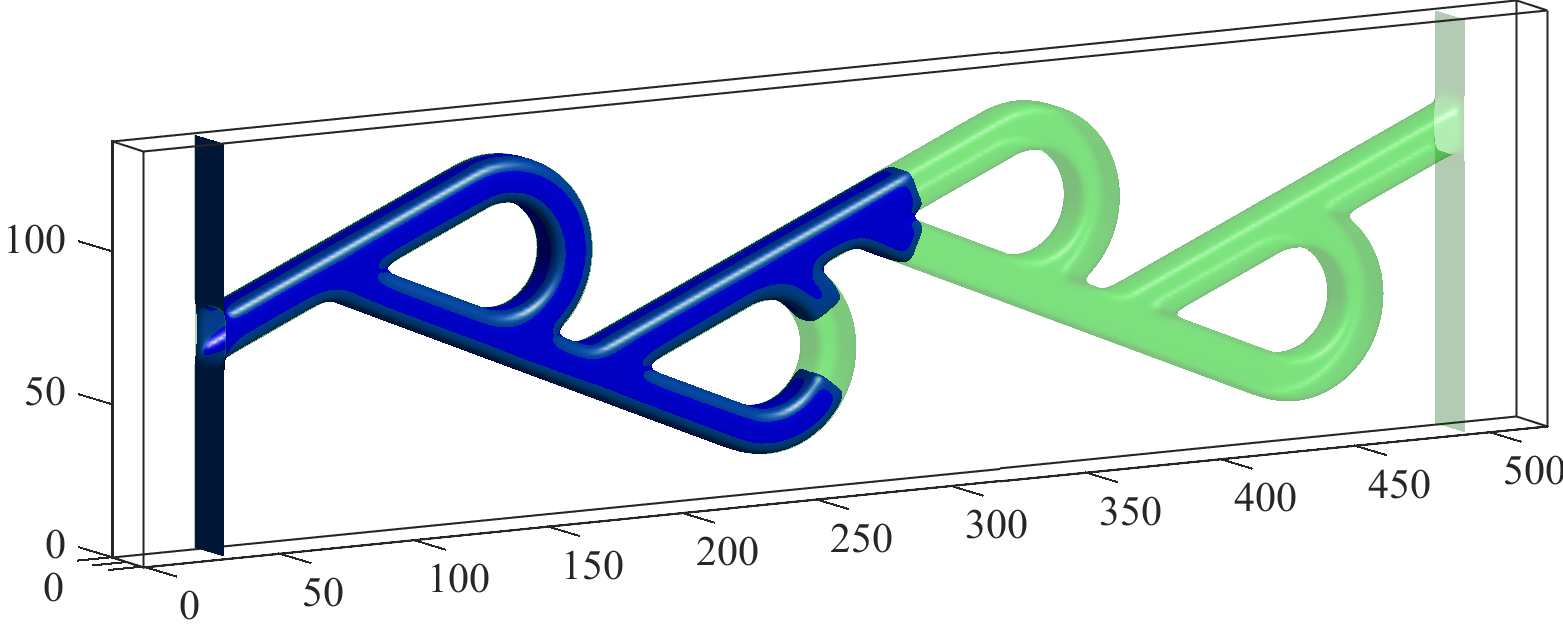}
	\end{minipage}}

	\subfigure[$t=1.5\times10^5$ \si{[ts]}]{
		\begin{minipage}{0.48\linewidth}
			\centering
			\includegraphics[width=3.0in]{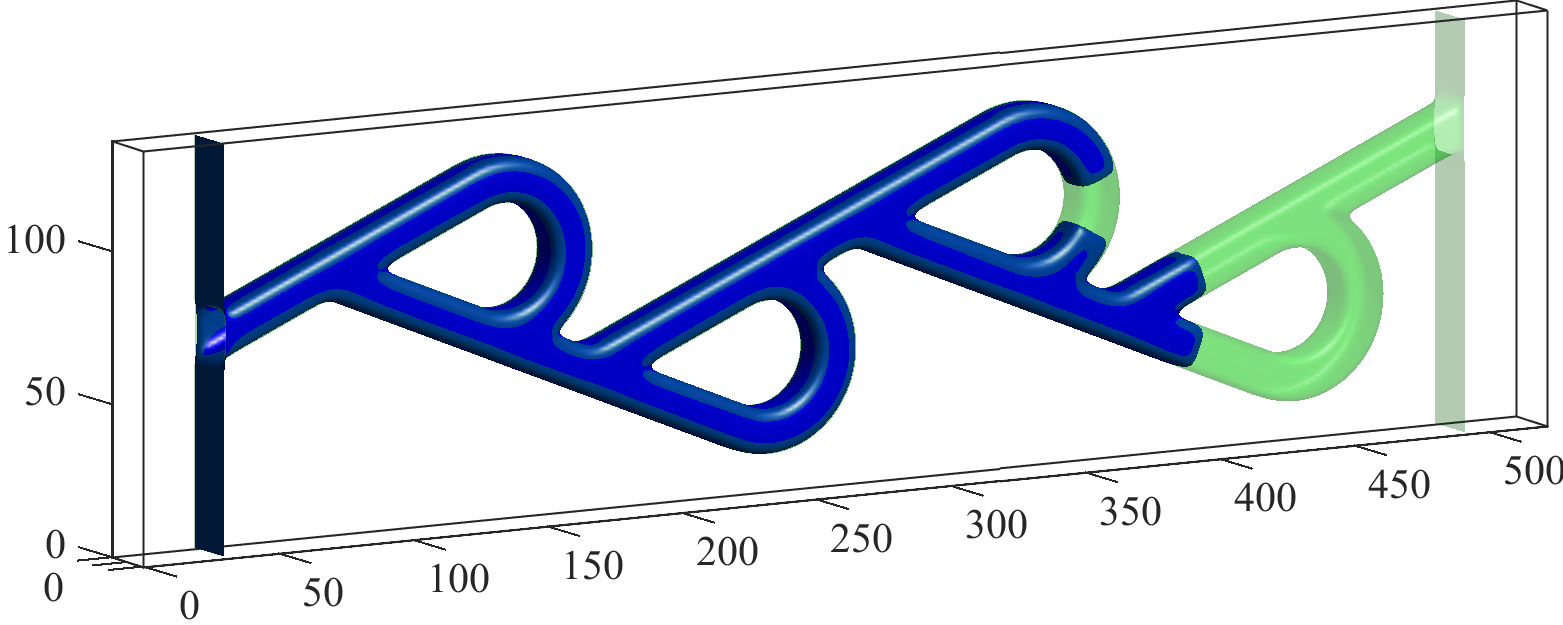}
	\end{minipage}}
	\subfigure[$t=2\times10^5$ \si{[ts]}]{
		\begin{minipage}{0.48\linewidth}
			\centering
			\includegraphics[width=3.0in]{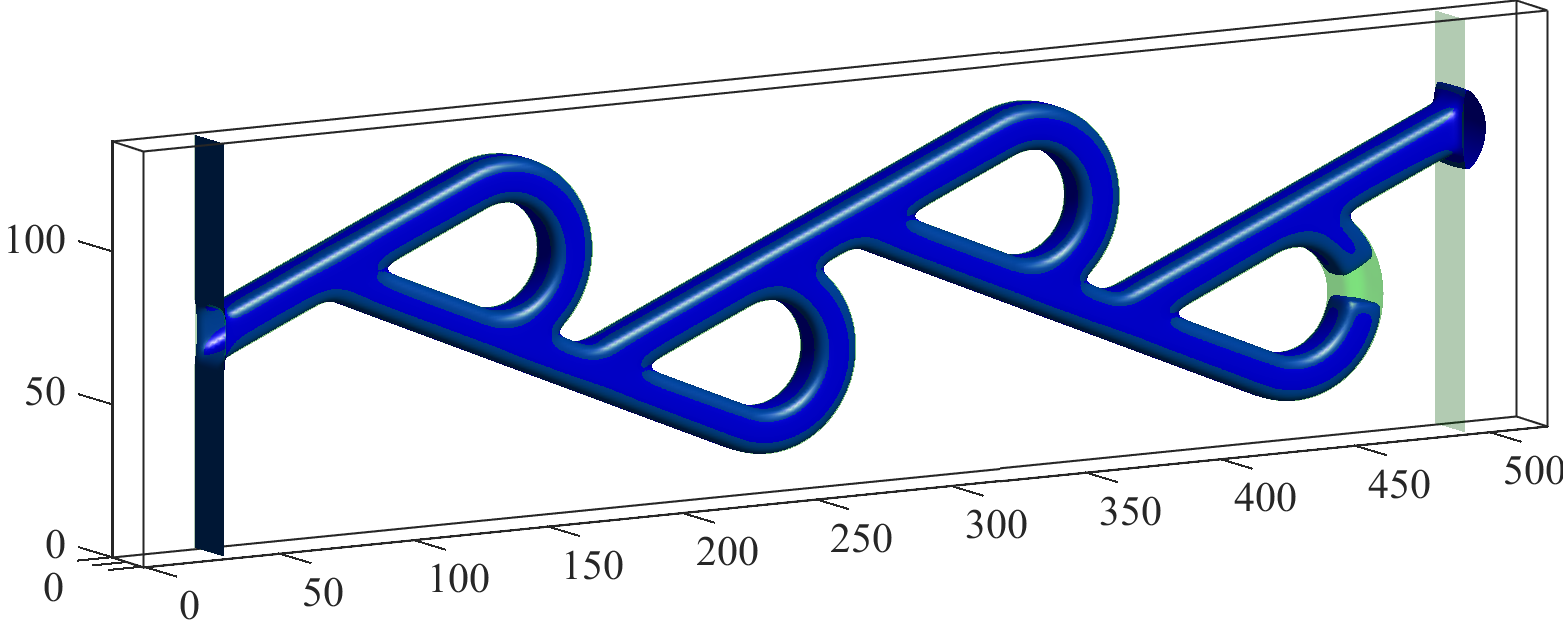}
	\end{minipage}}
	\caption{The snapshots of the reverse displacement in the 4 stages T45C Tesla valve.}
	\label{fig-Tesla3DReverse}
\end{figure}
\begin{figure}
	\centering
	\subfigure[Forward flow]{
		\begin{minipage}{0.2\linewidth}
			\centering
			\includegraphics[width=1.0in]{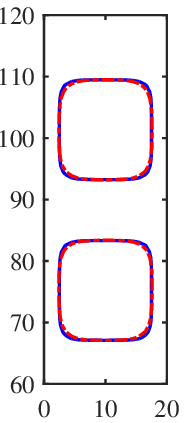}
	\end{minipage}}
	\subfigure[Reverse flow]{
		\begin{minipage}{0.2\linewidth}
			\centering
			\includegraphics[width=1.0in]{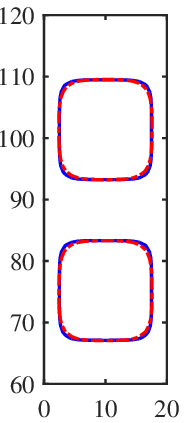}
		\end{minipage}}
	\caption{The profiles of the displacements at $x=300$ \si{[lu]} and $t=1.5\times10^5$ \si{[ts]}. Here the blue solid lines are the boundaries of pipes and red dotted lines represent the interfaces of two fluid phases.}
	\label{fig-Tesla3Dprofiles}
\end{figure}

Initially, the Tesla valve is full of the displaced fluid except for the driving fluid in the inlet section with the distance $d_s$ (see Fig. \ref{fig-Tesla3D}), and the fluids are driven by a pressure drop $\Delta P$ from inlet to outlet. We set $\Delta P=0.1$ \si{[mu/(lu.ts^2)]} for both forward and reverse two-phase displacements, and display the numerical results in Figs. \ref{fig-Tesla3DForward} and \ref{fig-Tesla3DReverse}. From these figures, one can observe that the driving fluid gradually fills the pipe under the action of the pressure drop in both forward and reverse cases. However, there are some differences in the middle process. In the forward case, when the driving fluid reaches to the bifurcation of the pipe, it tends to pass the straight pipe and the fluid in straight pipe goes faster than the another part in bend pipe. In contrast, in the reverse displacement, the fluids seem to have the same velocities in the two bifurcated pipes in each stage valve, and the driving fluid has the same tip positions in the two bifurcated pipes in $x$-direction. Additionally, one can also find that the trapped displaced fluid at the early process of the displacement disappears at the later process, which seems contrary to common sense. To explain this phenomenon, we plot the vertical profile of the pipe at $x=300$ \si{[lu]} and $t=1.5\times10^5$ \si{[ts]} in Fig. \ref{fig-Tesla3Dprofiles}. From this figure, one can observe that the driving fluid is not filled with the square profile of the pipe due to the larger inertia force caused by the pressure drop. As a result, the displaced fluid cannot be completely trapped, and the trapped parts can be displaced from four corners as time goes on.   

\begin{figure}
	\centering
	\includegraphics[width=3.5in]{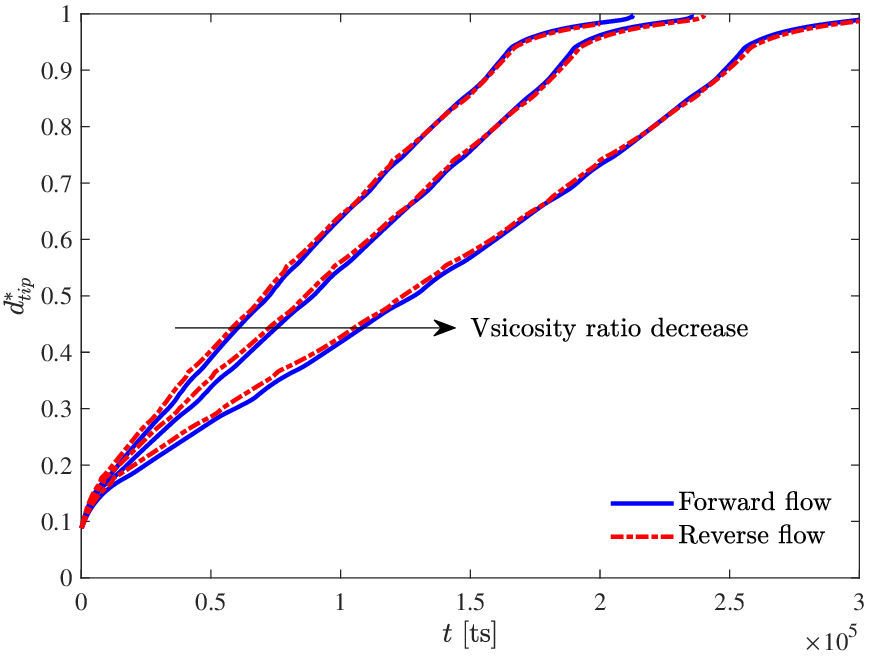}
	\caption{The normalized distance of the tip of the driving fluid from the inlet.}
	\label{fig-Tesla-Mu}
\end{figure}

The influence of the viscosity ratio ($M_{\mu}=\mu_1/\mu_2$) is further investigated, and $\mu_1$ is fixed at 0.1. We focus on two cases with $M_{\mu}=0.2,1$ and 5, and show the results of both forward and reverse displacements in Fig. \ref{fig-Tesla-Mu}. As shown in this figure, the normalized distances of the tip of the driving fluid from inlet in forward and reverse cases are basically the same as each other under the present pressure drop. 
In addition, with the decrease of $M_{\mu}$, the displaced fluid is more difficult to be displaced due to a larger viscous resistance, and thus the value of the distance $d_{tip}^*$ takes more time to reach 1.

\subsection{Displacement in a porous medium}\label{porous}
\begin{figure}
	\centering
	\subfigure[]{
		\begin{minipage}{0.32\linewidth}
			\centering
			\includegraphics[width=2.1in]{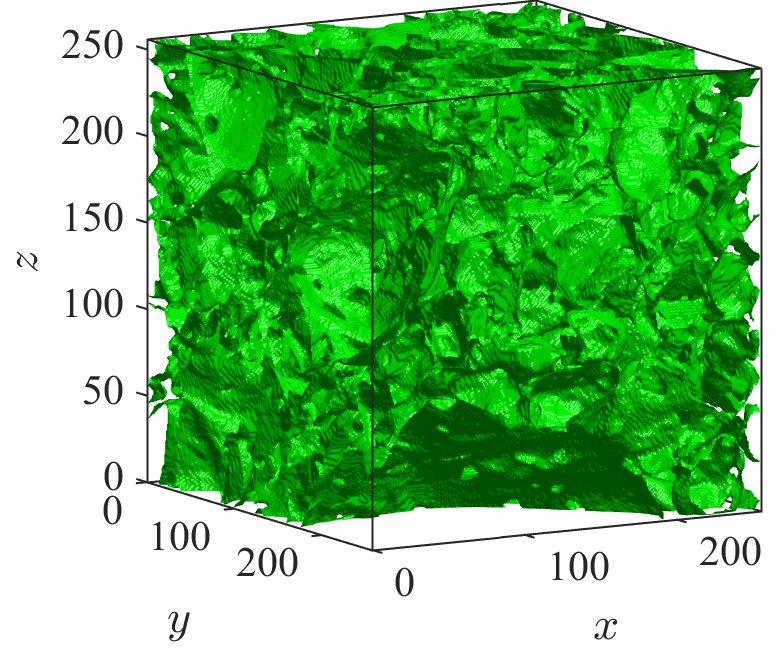}
	\end{minipage}}
	\subfigure[]{
		\begin{minipage}{0.32\linewidth}
			\centering
			\includegraphics[width=2.1in]{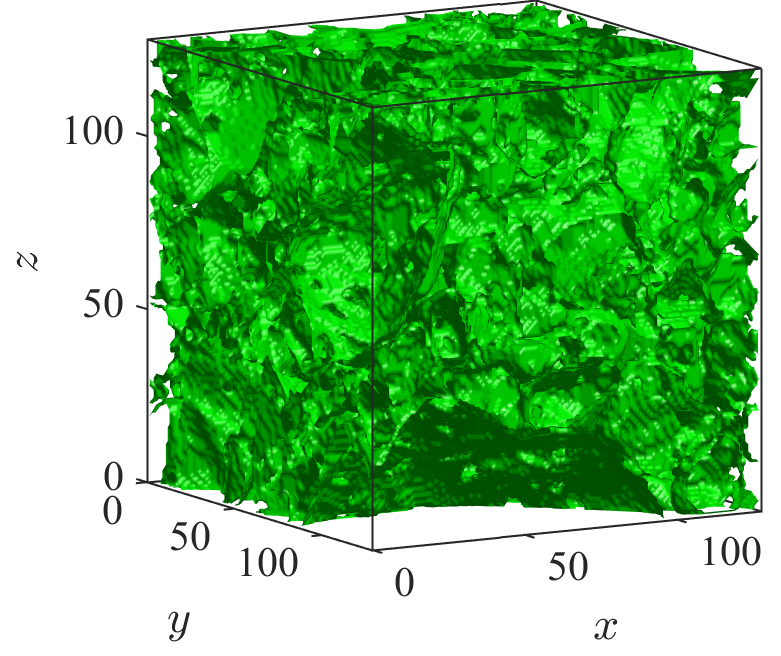}
	\end{minipage}}
	\subfigure[]{
		\begin{minipage}{0.32\linewidth}
			\centering
			\includegraphics[width=2.1in]{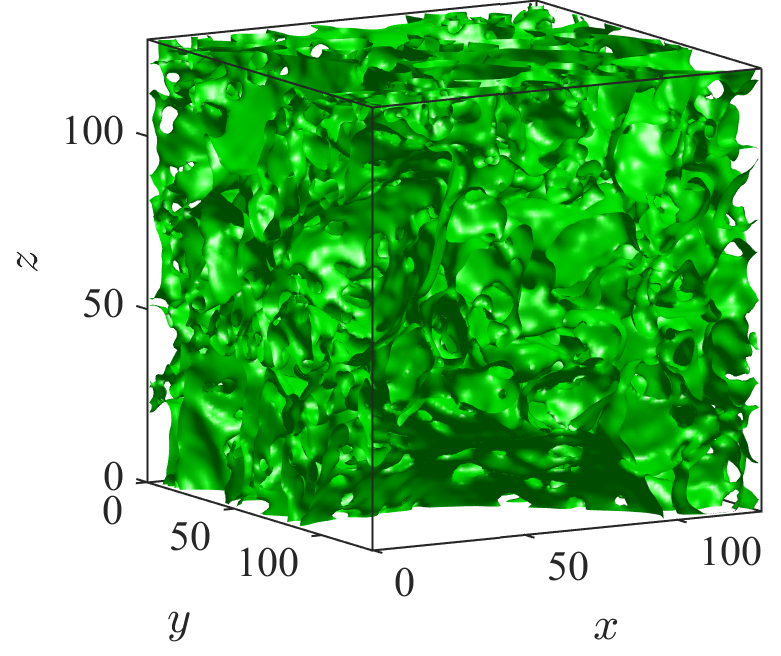}
	\end{minipage}}
	\caption{A sample of the poorly sorted unconsolidated fluvial sandpack [(a) The original $256\times256\times256$ micro-CT image with porosity $\varepsilon_o=0.3585$, (b) the compressed $128\times128\times128$ data with porosity $\varepsilon_c=0.3024$, (c) the smoothed $128\times128\times128$ data with porosity $\varepsilon_s=0.3068$].}
	\label{fig-porous-structure}
\end{figure}

We finally consider the displacement in a 3D porous medium, and the properties of fluids are the same as those in Section \ref{Tesla}. Figure \ref{fig-porous-structure}(a) shows a sample of the poorly sorted unconsolidated fluvial sandpack with a porosity $\varepsilon_o=0.3585$ \cite{Santos2022DB}, and the original $256\times256\times256$ data is firstly compressed into the size of $128\times128\times128$ \si{[lu^3]} to adapt to the memory limitation of the computing platform, as shown in Fig. \ref{fig-porous-structure}(b). To apply the present diffuse-interface method, the compressed data needs to be smoothed by the finite-time evolution of a standard CH equation for two phases, and Fig. \ref{fig-porous-structure}(c) plots the smoothed data after 200 iterations. In addition, the periodic boundary condition with a pressure drop \cite{Kim2007PF} is imposed on the inlet and outlet boundaries in $x$-direction and no-flux boundary condition is applied on other boundaries, which preserves the mass conservation of the system.  

The fluids are driven by a pressure drop $\Delta p=5\times10^{-3}$ \si{[mu/(lu.ts^2)]}, and some simulations are carried out to predict the relative permeabilities of the two-phase fluids in the porous medium. 
When the flows reach the steady state, the relative permeability of a specific fluid can be calculated by
\begin{equation}\label{eq-Kp}
	K_p=\frac{\int_{\bar{\Omega}}u_{x,p}\mathrm{d}\bar{\Omega}}{\int_{\bar{\Omega}}u'_{x,p}\mathrm{d}\bar{\Omega}},\quad p=1,2,
\end{equation}
where $\bar{\Omega}$ is the pore spaces ($\phi_0\leq0.5$) of the porous medium, $u_{x,p}$ and $u'_{x,p}$ are the $x$-component velocities of fluid $p$ in two-phase flow and single-phase flow, respectively. 

\begin{figure}
	\centering
	\includegraphics[width=2.1in]{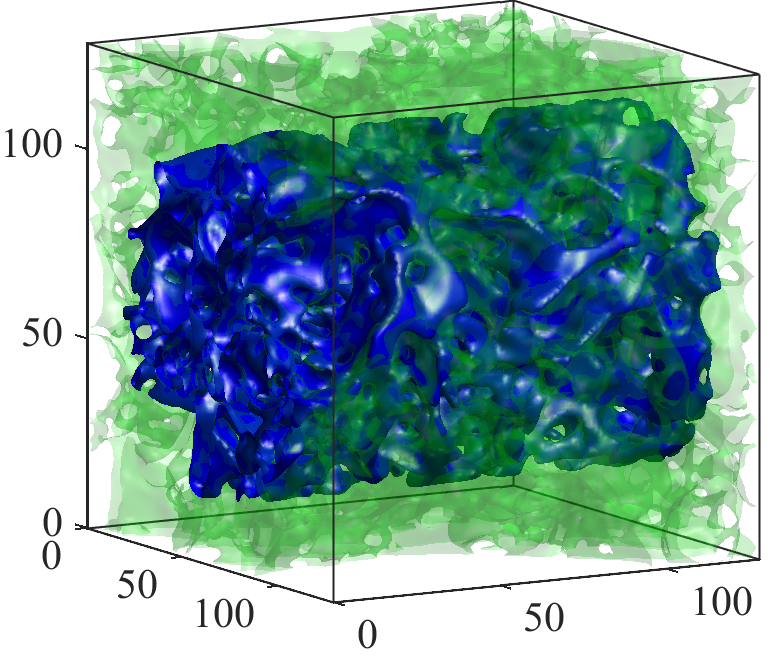}
	\caption{A example of the initial distribution of wetting fluid with $R_0=48$ \si{[lu]}.}
	\label{fig-porous-initial}
\end{figure}
\begin{figure}
	\centering
	\subfigure[$S_1=0.1668$]{
		\begin{minipage}{0.32\linewidth}
			\centering
			\includegraphics[width=2.1in]{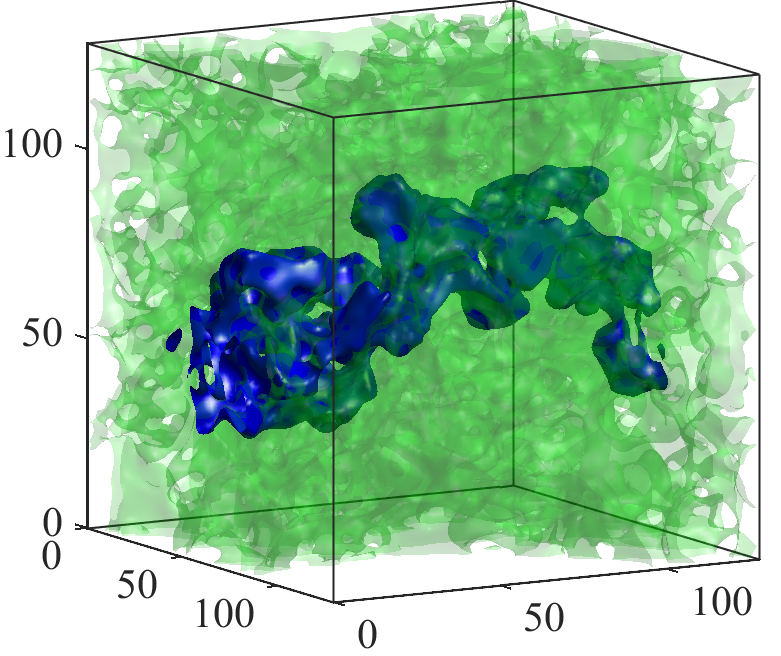}
	\end{minipage}}
	\subfigure[$S_1=0.2160$]{
		\begin{minipage}{0.32\linewidth}
			\centering
			\includegraphics[width=2.1in]{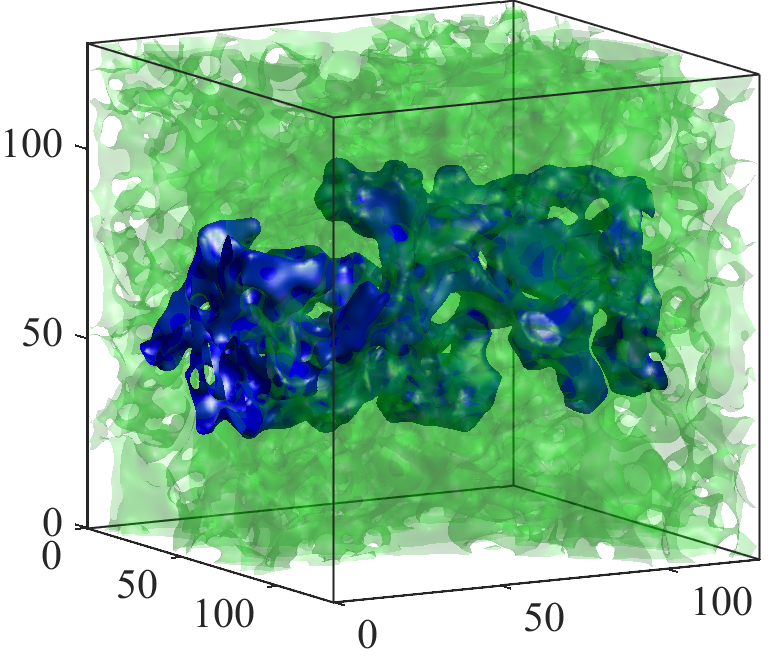}
	\end{minipage}}
	\subfigure[$S_1=0.2690$]{
		\begin{minipage}{0.32\linewidth}
			\centering
			\includegraphics[width=2.1in]{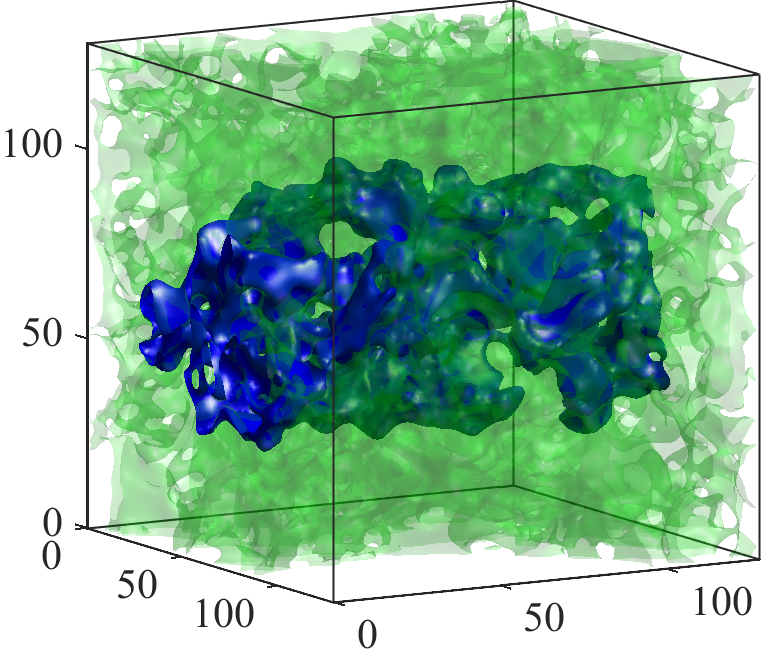}
	\end{minipage}}
	
	\subfigure[$S_1=0.3908$]{
		\begin{minipage}{0.32\linewidth}
			\centering
			\includegraphics[width=2.1in]{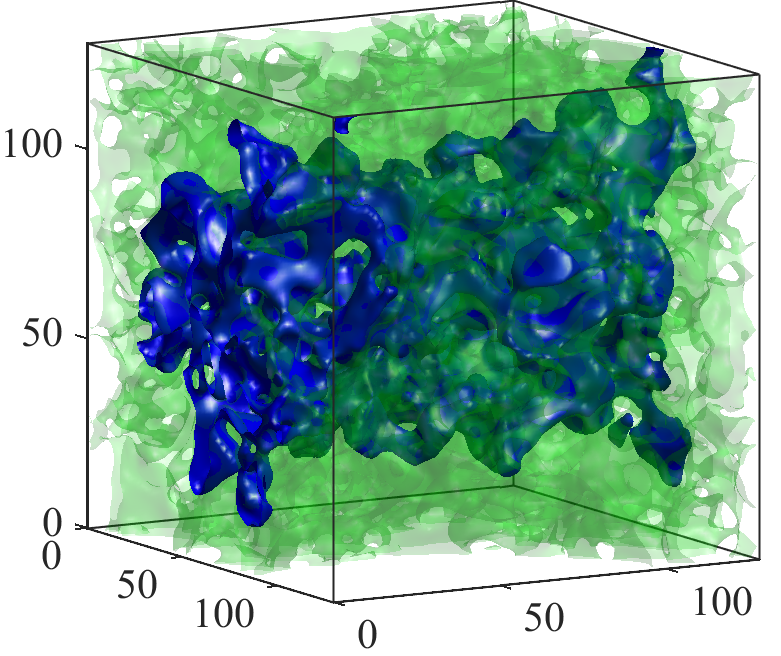}
	\end{minipage}}
	\subfigure[$S_1=0.5367$]{
		\begin{minipage}{0.32\linewidth}
			\centering
			\includegraphics[width=2.1in]{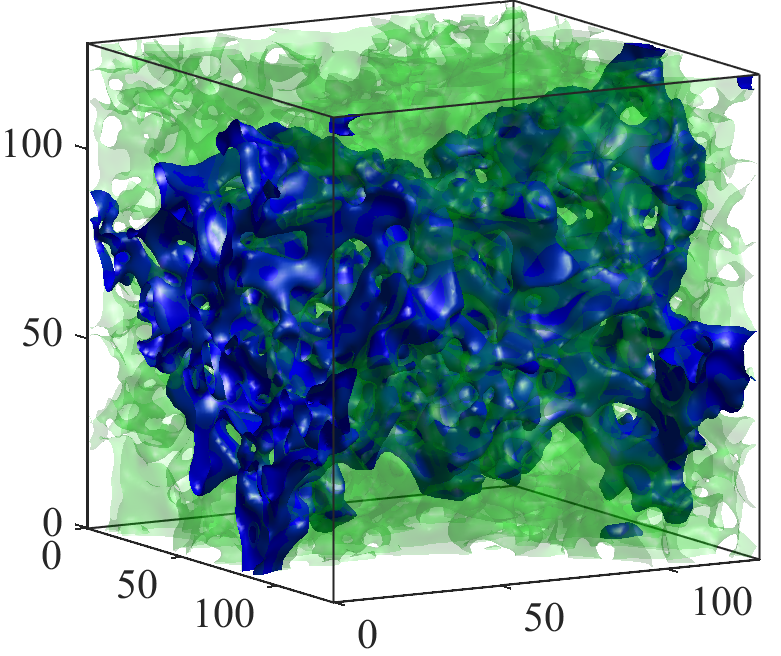}
	\end{minipage}}
	\subfigure[$S_1=0.7128$]{
		\begin{minipage}{0.32\linewidth}
			\centering
			\includegraphics[width=2.1in]{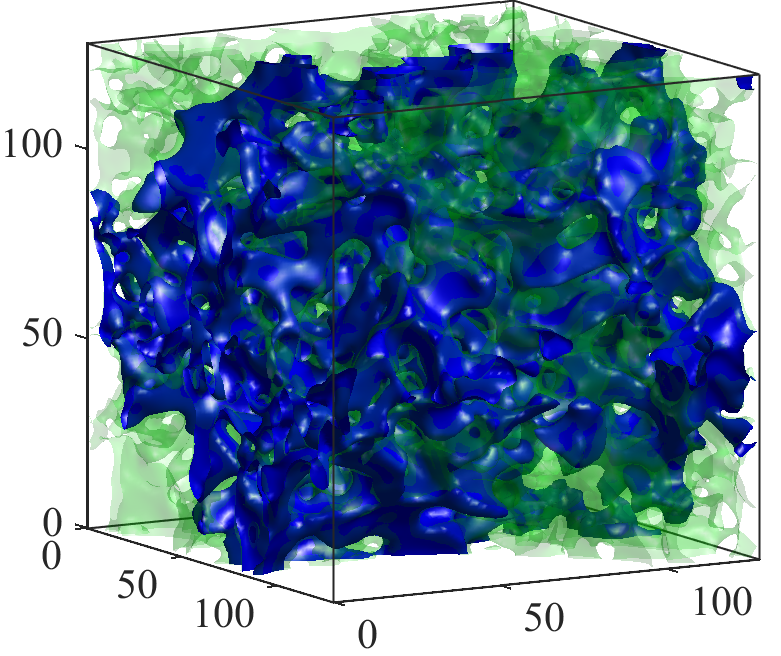}
	\end{minipage}}
	\caption{The finial distributions of wetting fluid in the porous medium at different saturations.}
	\label{fig-porous-Sw}
\end{figure} 
\begin{figure}
	\centering
	\includegraphics[width=3.5in]{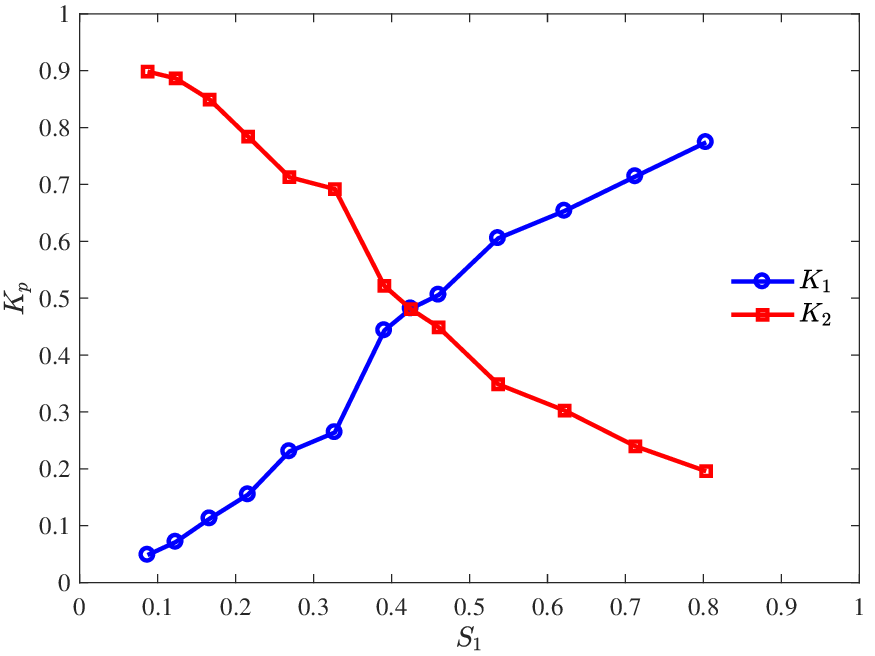}
	\caption{The relative permeabilities of wetting fluid ($K_1$) and non-wetting fluid ($K_2$) in the porous medium at $\theta=60^\circ$.}
	\label{fig-porous-KK}
\end{figure}
\begin{figure}
	\centering
	\includegraphics[width=3.5in]{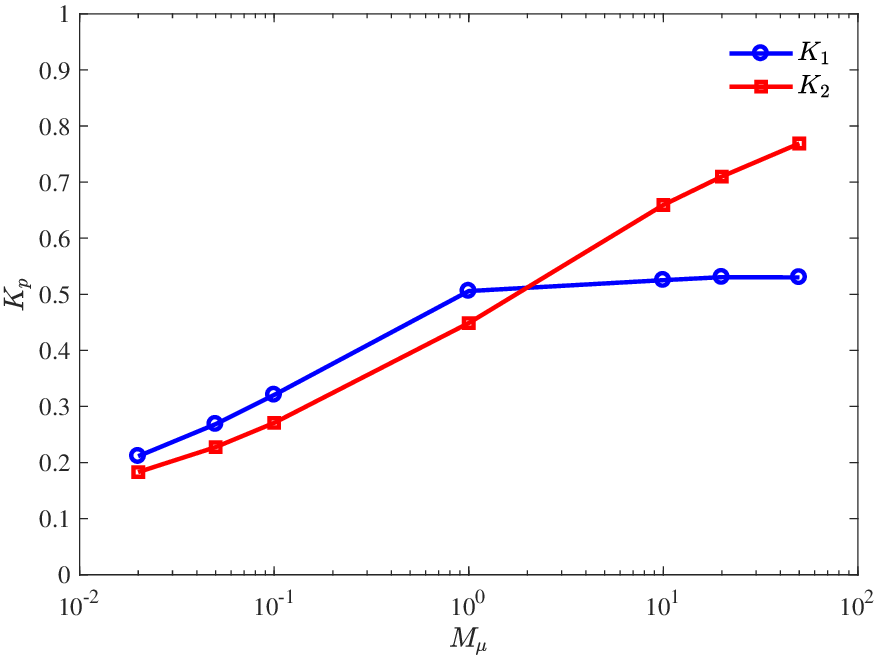}
	\caption{The relative permeabilities of wetting fluid ($K_1$) and non-wetting fluid ($K_2$) in the porous medium at $S_1=0.4605$.}
	\label{fig-porous-mu}
\end{figure}
\begin{figure}
	\subfigure[$M_{\mu}=0.02$]{
		\begin{minipage}{0.32\linewidth}
			\centering
			\includegraphics[width=2.1in]{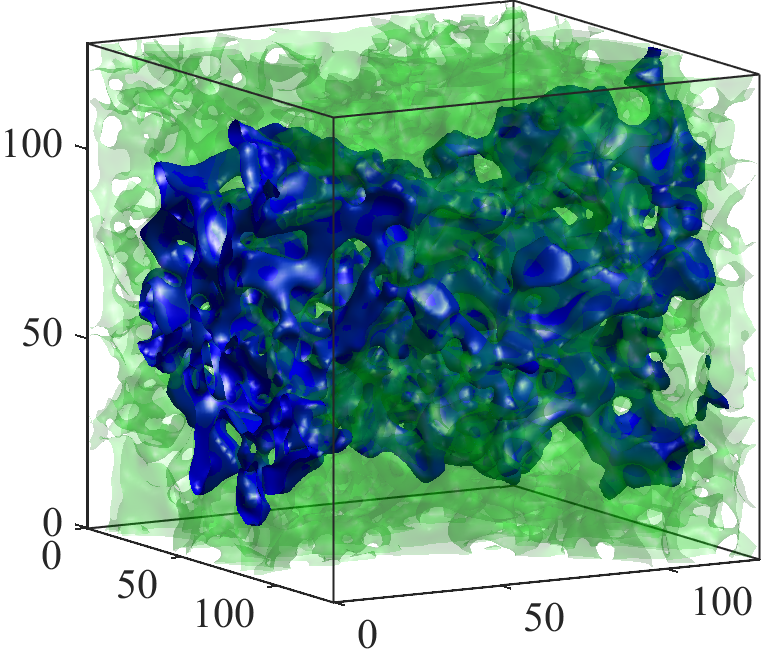}
	\end{minipage}}
	\subfigure[$M_{\mu}=1$]{
		\begin{minipage}{0.32\linewidth}
			\centering
			\includegraphics[width=2.1in]{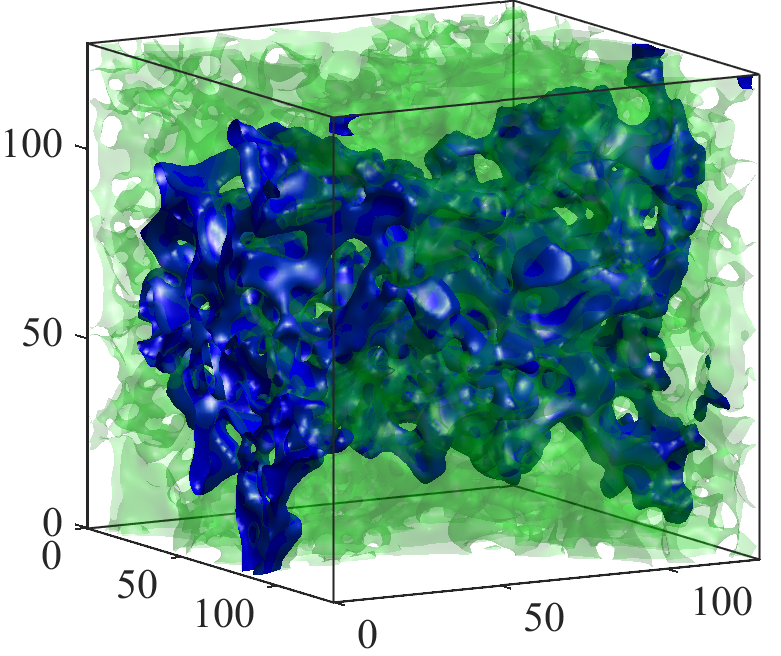}
	\end{minipage}}
	\subfigure[$M_{\mu}=50$]{
		\begin{minipage}{0.32\linewidth}
			\centering
			\includegraphics[width=2.1in]{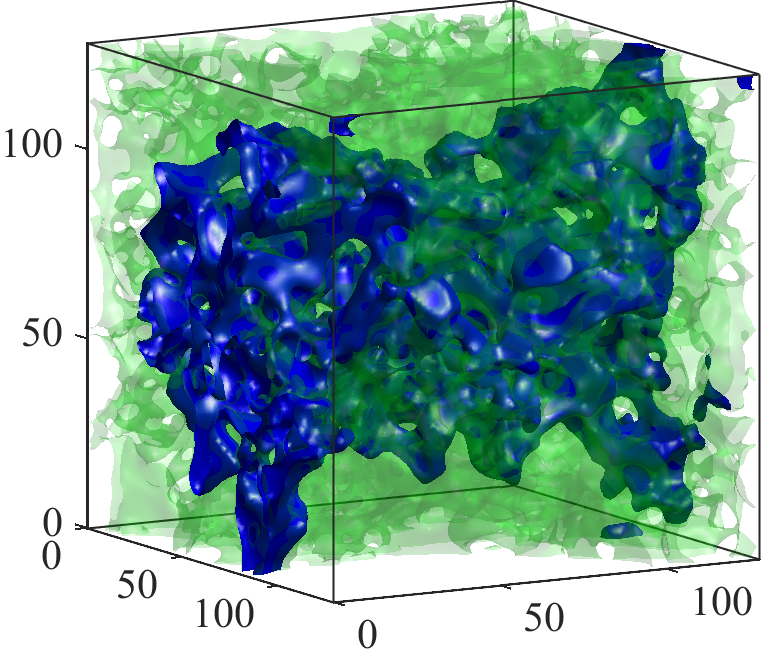}
	\end{minipage}}
	\caption{The finial distributions of wetting fluid in the porous medium with different viscosity ratios at $S_1=0.4605$.}
	\label{fig-porous-Mu}
\end{figure}
\begin{figure}
	\centering
	\includegraphics[width=3.5in]{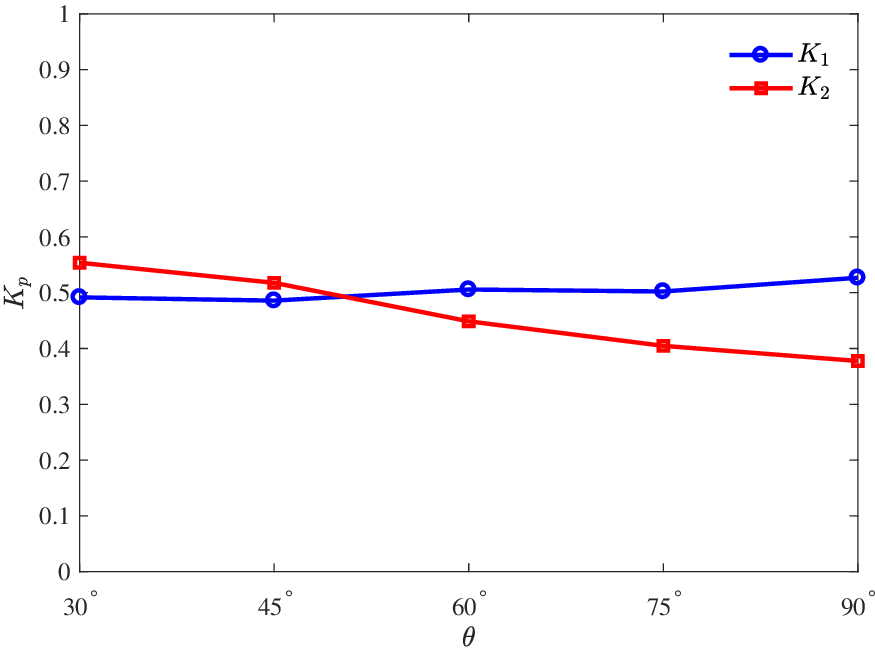}
	\caption{The relative permeabilities of wetting fluid ($K_1$) and non-wetting fluid ($K_2$) in the porous medium at $S_1=0.4605$.}
	\label{fig-porous-theta}
\end{figure}
\begin{figure}
	\subfigure[$\theta=30^\circ$]{
		\begin{minipage}{0.32\linewidth}
			\centering
			\includegraphics[width=2.1in]{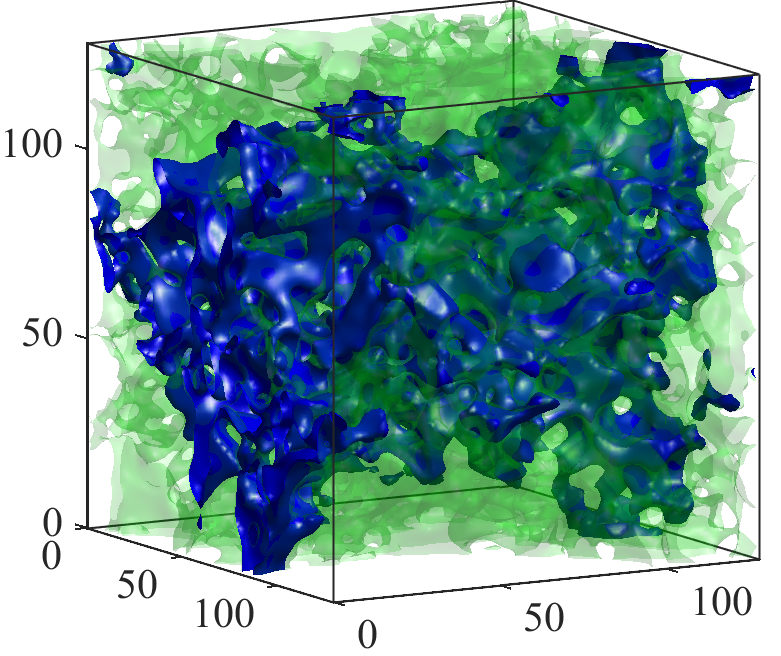}
	\end{minipage}}
	\subfigure[$\theta=60^\circ$]{
		\begin{minipage}{0.32\linewidth}
			\centering
			\includegraphics[width=2.1in]{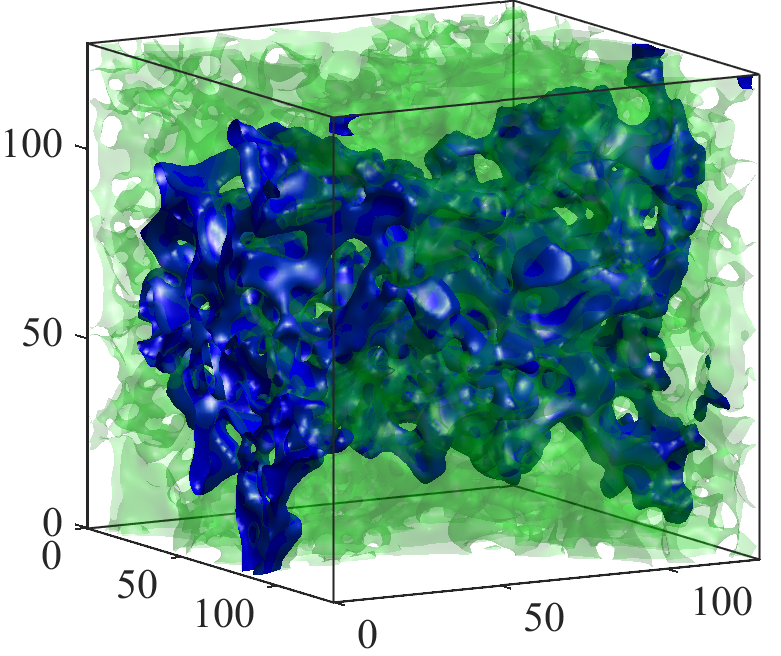}
	\end{minipage}}
	\subfigure[$\theta=90^\circ$]{
		\begin{minipage}{0.32\linewidth}
			\centering
			\includegraphics[width=2.1in]{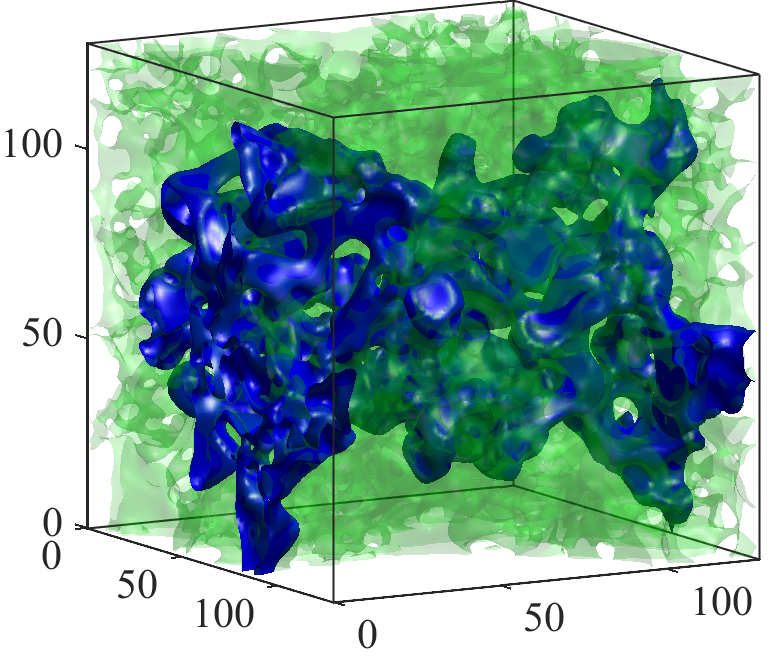}
	\end{minipage}}
	\caption{The finial distributions of wetting fluid in the porous medium with different contact angles at $S_1=0.4605$.}
	\label{fig-porous-Theta}
\end{figure}

We first investigate the effect of the saturation on the relative permeabilities under the condition of $\theta=60^\circ$. The saturation of wetting fluid $S_1$ is determined by the radius $R_0$ and the following initialization of $\phi_1$,
\begin{equation}
	\phi_1\left(x,y,z\right)= \left[1-\phi_0\left(x,y,z\right)\right]\left[\frac{1}{2}+\frac{1}{2}\tanh\frac{R_0-\sqrt{\left(y-Ly/2\right)^2+\left(z-Lz/2\right)^2}}{D/2}\right],
\end{equation}
and Fig. \ref{fig-porous-initial} shows an example of the initial distribution of wetting fluid with $R_0=48$ \si{[lu]}. After some simulations, we plot the finial distributions of wetting fluid under different saturations (or radii) in Fig. \ref{fig-porous-Sw}. From this figure, one can observe that when the saturation of wetting fluid is increased, the wetting fluid flows into more pores and the connectivity of wetting phase gradually increases. To quantitatively analyze the present results, we also calculate the relative permeabilities at different saturations, and plot them in Fig. \ref{fig-porous-KK}. As shown in this figure, the relative permeability of the wetting fluid increases with the increase of $S_1$, while that of the non-wetting fluid decreases. Actually, when the value of $S_1$ is small, as shown in Fig. \ref{fig-porous-Sw}(a), the wetting fluid has fewer pathways in $x$-direction, which causes a low Darcy velocity. As a result, the relative permeability of wetting fluid is small. In contrast, when $S_1$ is high, the wetting fluid fills into the most of the pores, and forms many flow paths from inlet to outlet such that the relative permeability of wetting fluid becomes large. In addition, when the wetting fluid saturation is $50\%$, the relative permeability of the wetting fluid is larger than that of non-wetting fluid because of the effect of the wettability and the internal distribution of wetting fluid in the porous medium. 

We then consider the influence of the viscosity ratio $M_{\mu}=\mu_1/\mu_2$ with $\mu_1=0.1$, and the saturation of the wetting fluid is set as $S_1=0.4605$ ($R_0=48$). As seen from Fig. \ref{fig-porous-mu}, when the viscosity ratio is varied from $M_{\mu}=0.02$ to $M_{\mu}=50$, the flow resistance exerted on the non-wetting fluid decreases, which results in the increase of the relative permeability $K_2$. However, the variation of the relative permeability $K_1$ shows some differences. When $M_{\mu}<1$, the viscosity of non-wetting fluid decreases with the increase of $M_{\mu}$, and thus the wetting fluid is easier to drive the non-wetting fluid and has a better connectivity [see the final shapes in Figs. \ref{fig-porous-Mu}(a) and (b)]. However, when $\mu_2<\mu_1$, the increase of $K_1$ is small because of the dominant flow formed by the non-wetting fluid [see Figs. \ref{fig-porous-Mu}(b) and (c)].           

The effect of the wettability is also discussed at $S_1=0.4605$. Figure \ref{fig-porous-theta} plots the relative permeability $K_p$ as a function of the contact angle. From this figure, one can find that the contact angle has a significant effect on the relative permeabilities. With the increase of the contact angle, the relative permeability of wetting fluid increases a little, while that of non-wetting fluid decreases a lot, this is because of the weakening of the adhesive force between the wetting fluid and solid surface and the strengthening of the adhesive force between the non-wetting fluid and solid surface. We also show the distributions of wetting fluid at different contact angles in Fig. \ref{fig-porous-Theta}, and find that when the solid surface become more hydrophilic, the wetting fluid prefers to flow into small pores to minimize the surface energy.

\section{Conclusions}\label{Conclusions}
In this paper, we first proposed a ternary phase-field model for two-phase flows in complex geometries. In this model, a specific expression of the liquid-solid surface tension coefficient is designed, and it is assumed to be related to the curvature and the contact angle of the solid surface. Then the reformulated consistent NS equations is used to describe the fluid flows, in which the phase-field variable for solid phase is introduced to achieve a high viscosity in solid phase and preserve the velocity boundary conditions on solid surface. To solve the present CH and NS equations for two-phase flows in complex geometries, the Hermite-moment based LB method is further developed, in which the collision step is conducted in the natural moment space with a lower triangular relaxation matrix, and an adjustable scale factor is also introduced to improve the numerical stability. To test the present MRT-LB method, several benchmark problems are considered, and the numerical results are in good agreement with the analytical solutions. Finally, the present model is applied to study two 3D displacement problems in a complex channel and a porous medium, and the simulation results illustrate the capacity of the present model in the study of two-phase flows in complex geometries. In the future, the present model will be extended to investigate the problems of two-phase flows with moving solid objects or solid-liquid phase transition.

\section*{Acknowledgments}
This research has been supported by the National Natural Science Foundation of China under Grants No. 12072127 and No 51836003, and the Interdiciplinary Research Program of HUST (2023JCYJ002). The computation was completed on the HPC Platform of Huazhong University of Science and Technology.

\appendix
\section{The weight coefficients and transformation matrices of D2Q9 and Q3Q15 lattice structures}\label{Matrices}
In this appendix, we list the related weight coefficients and transformation matrices of the D2Q9 and D3Q15 lattice structures. Here we set $d_0^p=d_0$ and $\omega_{i,p}=\omega_i$ for simplicity, and $d_0=0.5$ is used in our numerical simulations.

\noindent D2Q9: $\omega_0=1-2d_0+d_0^2$, $\omega_{1-4}=\left(d_0-d_0^2\right)/2$, and $\omega_{5-8}=d_0^2/4$. $\mathbf{C}_d=\mathbf{diag}\left(1,c,c,c^2,c^2,c^2,c^3,c^3,c^4\right)$, $\mathbf{K}=\mathbf{diag}\left(s_0,s_1,s_1,s_{2a},s_{2a},s_{2b},s_3,s_3,s_4\right)$,
\begin{equation}
	\mathbf{M}_0=\begin{pmatrix*}[r]
		1 &  1 &  1 &  1 &  1 &  1 &  1 &  1 &  1\\
		0 &  1 &  0 & -1 &  0 &  1 & -1 & -1 &  1\\
		0 &  0 &  1 &  0 & -1 &  1 &  1 & -1 & -1\\
		0 &  1 &  0 &  1 &  0 &  1 &  1 &  1 &  1\\
		0 &  0 &  1 &  0 &  1 &  1 &  1 &  1 &  1\\		
		0 &  0 &  0 &  0 &  0 &  1 & -1 &  1 & -1\\	
		0 &  0 &  0 &  0 &  0 &  1 & -1 & -1 &  1\\	
		0 &  0 &  0 &  0 &  0 &  1 &  1 & -1 & -1\\
		0 &  0 &  0 &  0 &  0 &  1 &  1 &  1 &  1\\
	\end{pmatrix*},\quad
	\mathbf{N}=\begin{pmatrix*}[r]
		1 & 0 & 0 & 0 & 0 & 0 & 0 & 0 & 0 \\
		0 & 1 & 0 & 0 & 0 & 0 & 0 & 0 & 0 \\
		0 & 0 & 1 & 0 & 0 & 0 & 0 & 0 & 0 \\
		-c_s^2 & 0 & 0 & 1 & 0 & 0 & 0 & 0 & 0 \\
		-c_s^2 & 0 & 0 & 0 & 1 & 0 & 0 & 0 & 0 \\
		0 & 0 & 0 & 0 & 0 & 1 & 0 & 0 & 0 \\
		0 & -c_s^2 & 0 & 0 & 0 & 0 & 1 & 0 & 0 \\
		0 & 0 & -c_s^2 & 0 & 0 & 0 & 0 & 1 & 0 \\
		c_s^4 & 0 & 0 & -c_s^2 & -c_s^2 & 0 & 0 & 0 & 1 \\
	\end{pmatrix*}.
\end{equation}

\noindent D3Q15: $\omega_0=1-3d_0+2d_0^2$, $\omega_{1-6}=\left(d_0-d_0^2\right)/2$, and $\omega_{7-14}=d_0^2/8$. $\mathbf{C}_d=\mathbf{diag}\left(1,c,c,c,c^2,c^2,c^2,c^2,c^2,c^2,c^3,c^3,c^3,c^3,c^4\right)$, $\mathbf{K}=\mathbf{diag}\left(s_0,s_1,s_1,s_1,s_{2a},s_{2a},s_{2a},s_{2b},s_{2b},s_{2b},s_3,s_3,s_3,s_3,s_4\right)$,
\begin{equation}
	\setcounter{MaxMatrixCols}{15}
	\mathbf{M}_0=\begin{pmatrix*}[r]
		1 &  1 &  1 &  1 &  1 &  1 &  1 &  1 &  1 &  1 &  1 &  1 &  1 &  1 &  1\\
		0 &  1 & -1 &  0 &  0 &  0 &  0 &  1 & -1 &  1 & -1 &  1 & -1 &  1 & -1\\
		0 &  0 &  0 &  1 & -1 &  0 &  0 &  1 &  1 & -1 & -1 &  1 &  1 & -1 & -1\\
		0 &  0 &  0 &  0 &  0 &  1 & -1 &  1 &  1 &  1 &  1 & -1 & -1 & -1 & -1\\
		0 &  1 &  1 &  0 &  0 &  0 &  0 &  1 &  1 &  1 &  1 &  1 &  1 &  1 &  1\\
		0 &  0 &  0 &  1 &  1 &  0 &  0 &  1 &  1 &  1 &  1 &  1 &  1 &  1 &  1\\
		0 &  0 &  0 &  0 &  0 &  1 &  1 &  1 &  1 &  1 &  1 &  1 &  1 &  1 &  1\\
		0 &  0 &  0 &  0 &  0 &  0 &  0 &  1 & -1 & -1 &  1 &  1 & -1 & -1 &  1\\
		0 &  0 &  0 &  0 &  0 &  0 &  0 &  1 &  1 & -1 & -1 & -1 & -1 &  1 &  1\\
		0 &  0 &  0 &  0 &  0 &  0 &  0 &  1 & -1 &  1 & -1 & -1 &  1 & -1 &  1\\
		0 &  0 &  0 &  0 &  0 &  0 &  0 &  1 & -1 &  1 & -1 &  1 & -1 &  1 & -1\\
		0 &  0 &  0 &  0 &  0 &  0 &  0 &  1 &  1 & -1 & -1 &  1 &  1 & -1 & -1\\
		0 &  0 &  0 &  0 &  0 &  0 &  0 &  1 &  1 &  1 &  1 & -1 & -1 & -1 & -1\\
		0 &  0 &  0 &  0 &  0 &  0 &  0 &  1 & -1 & -1 &  1 & -1 &  1 &  1 & -1\\
		0 &  0 &  0 &  0 &  0 &  0 &  0 &  1 &  1 &  1 &  1 &  1 &  1 &  1 &  1\\
	\end{pmatrix*},
\end{equation}
\begin{equation}
	\setcounter{MaxMatrixCols}{15}
	\mathbf{N}_H=\begin{pmatrix*}[r]
		1 & 0 & 0 & 0 & 0 & 0 & 0 & 0 & 0 & 0 & 0 & 0 & 0 & 0 & 0\\
		0 & 1 & 0 & 0 & 0 & 0 & 0 & 0 & 0 & 0 & 0 & 0 & 0 & 0 & 0\\
		0 & 0 & 1 & 0 & 0 & 0 & 0 & 0 & 0 & 0 & 0 & 0 & 0 & 0 & 0\\
		0 & 0 & 0 & 1 & 0 & 0 & 0 & 0 & 0 & 0 & 0 & 0 & 0 & 0 & 0\\
		-c_s^2 & 0 & 0 & 0 & 1 & 0 & 0 & 0 & 0 & 0 & 0 & 0 & 0 & 0 & 0\\
		-c_s^2 & 0 & 0 & 0 & 0 & 1 & 0 & 0 & 0 & 0 & 0 & 0 & 0 & 0 & 0\\
		-c_s^2 & 0 & 0 & 0 & 0 & 0 & 1 & 0 & 0 & 0 & 0 & 0 & 0 & 0 & 0\\
		0 & 0 & 0 & 0 & 0 & 0 & 0 & 1 & 0 & 0 & 0 & 0 & 0 & 0 & 0\\
		0 & 0 & 0 & 0 & 0 & 0 & 0 & 0 & 1 & 0 & 0 & 0 & 0 & 0 & 0\\
		0 & 0 & 0 & 0 & 0 & 0 & 0 & 0 & 0 & 1 & 0 & 0 & 0 & 0 & 0\\
		0 & -c_s^2 & 0 & 0 & 0 & 0 & 0 & 0 & 0 & 0 & 1 & 0 & 0 & 0 & 0\\
		0 & 0 & -c_s^2 & 0 & 0 & 0 & 0 & 0 & 0 & 0 & 0 & 1 & 0 & 0 & 0\\
		0 & 0 & 0 & -c_s^2 & 0 & 0 & 0 & 0 & 0 & 0 & 0 & 0 & 1 & 0 & 0\\
		0 & 0 & 0 & 0 & 0 & 0 & 0 & 0 & 0 & 0 & 0 & 0 & 0 & 1 & 0\\
		2c_s^4 & 0 & 0 & 0 & -c_s^2 & -c_s^2 & -c_s^2 & 0 & 0 & 0 & 0 & 0 & 0 & 0 & 1\\
	\end{pmatrix*}.
\end{equation} 
Here it should be noted that in the D3Q15 lattice structure, the Hermite matrix $\mathbf{H}=\mathbf{N}_H\mathbf{M}$ cannot be directly obtained from the commonly used Hermite polynomials, and some row transformations for the Hermite moments are needed to satisfy the weighted orthogonality of $\mathbf{H}$.
\bibliographystyle{elsarticle-num} 
\bibliography{references}
\end{document}